\documentclass[useAMS,usenatbib,usegraphicx]{mn2e}
\usepackage{times}
\usepackage{amsmath}

\newcommand{\msun}{\,\mbox{$\rm M_{\odot}$}}
\newcommand{\mearth}{\,\mbox{$\rm M_{\oplus}$}}

\newcommand{\kms}{\hbox{km s$^{-1}$}}
\newcommand{\ms}{\hbox{m s$^{-1}$}}

\newcommand{\vsini}{\hbox{$v$\,sin\,$i$}}

\newcommand{\degs}{$\degr$}

\newcommand{\ha}{H$\alpha$}

\newcommand{\tps}{\hbox{$T_{p}/T_{s}$}}

\title[Effect of Spots on Radial Velocity Curves]{The effect of M dwarf
starspot activity on low-mass planet detection thresholds\textbf{} }
\author[J.R.~Barnes and S.V.~Jeffers and H.R.A.~Jones]{J.R.~Barnes$^{1}$, and
S.V.~Jeffers$^{2}$, and H.R.A.~Jones$^{1}$\\
$^{1}$ Centre for Astrophysics Research,. University of Hertfordshire,. College Lane, Hatfield. Herts. AL10 9AB. UK \\
$^{2}$ University of Utrecht, P.O. Box 80000, 3508 TA, Utrecht, The Netherlands}

\begin{document}

\date{Accepted for publication in MNRAS, 2010 November 4. Received 2010 November 3; in original form from 2010 March 19.}

\pagerange{\pageref{firstpage}--\pageref{lastpage}} \pubyear{2010}

\maketitle

\protect\label{firstpage}

\begin{abstract}
In light of the growing interest in searching for low mass, rocky planets, we
investigate the impact of starspots on radial velocity searches
for earth-mass planets in orbit about M dwarf stars. Since new surveys
targeting M dwarfs will likely be carried out at infrared wavelengths, a
comparison between {V and Y band starspot induced jitter is made, indicating
a reduction of up to an order of magnitude when observing in the Y band. The
exact reduction in jitter is dependent on the photosphere to spot contrast
ratio, with greater improvements at smaller contrasts.

We extrapolate a model used to describe solar spot distributions to
simulate the spot patterns that we expect to find on M dwarfs. Under the
assumption that M dwarfs are near or fully convective, we randomly place
starspots on the stellar surface, simulating different levels of spot coverage.
Line profiles, distorted by spots are derived and are used to investigate the
starspot induced jitter. By making assumptions about the degree of spot
activity, detection limits for earth-mass planets in habitable zones are
simulated for between 10 and 500 observation epochs. We find that $\leq$ 50
epochs are required to detect 1\,-\,2 \mearth\ planets (with $< 1$ per cent
false alarm probability) orbiting slowly rotating 0.1 and 0.2 \msun\ stars. This
sensitivity decreases when typical rotation velocities and activity levels for
each stellar mass/spectral type are considered. No detections of below 20
\mearth\ planets are expected for $\leq$ 500 observations for the most active
stars with \hbox{\vsini\ $\geq$ 20 \kms} and dark spots.
}
\end{abstract}


\section{Introduction}

Nearly 500 extrasolar planets in over 400 planetary systems have been
discovered\footnote{www.exoplanet.eu}, and yet $< 6$ per cent of those
with known or estimated host star masses are M dwarf planetary systems. Since M
dwarf masses may be as low as one tenth of a solar mass, it is not unreasonable
to expect that their most massive planets might also be correspondingly smaller
\citep{ida05} and thus more difficult to detect than the ubiquitous gas
giants found orbiting F, G \& K dwarf stars. { On the other hand, planets of a
fixed mass orbiting lower mass stars become easier to detect, owing to the reduced
mass ratio of the system. Detecting low-mass planets
(which are predicted in significant numbers by \cite{ida05}) is however no easy task,
since the planet-induced stellar velocity amplitudes are small, even for M dwarfs.}
Additionally, in the case of M dwarfs, there
is an observational bias against detecting planets at optical wavelengths
(where most searches have so far been carried out) owing to the relatively low
fluxes compared with redder wavelengths. Observations carried out at near infrared 
wavelengths, where the host M stars are several magnitudes brighter than at shorter 
wavelengths, open up the possibility of carrying out larger surveys of lower mass 
stars that are capable of detecting low-mass planets.

Radial velocity surveys are still by far the most successful means of detecting
planets. Despite the difficulties of observing in the infrared, there are a
number of fledgling projects aimed at carrying out planet searches. Both
\cite{ramsey08} and \cite{steinmetz08} showed that high precision ($< 10$ \ms)
radial velocity measurements could be made at short infrared wavelengths via
observations of the Sun. \cite{seifahrt08crires} demonstrated that similar
precision was possible on short timescales using the Cryogenic high-resolution
infrared \'{e}chelle spectrograph (CRIRES), operating at L-band wavelengths, at the
Very Large Telescope (VLT). More recently, an ammonia gas cell has been
employed to achieve $\sim 5$ \ms\ precision \citep{bean09crires} over six month
timescales with CRIRES operating at K-band wavelengths. This survey
\citep{bean09vb10} has not found evidence for the massive planet claimed to be
orbiting the low-mass M dwarf VB 10 \citep{pravdo09vb10,zapatero09vb10}.  

With infrared multi-order spectrographs working at high 
resolution \hbox{($R > 50,000$)},
precision radial velocities of order 1 \ms\ are expected to be achieved in the
next few years. Instrumental precision of 1 \hbox{\ms}\ is a desirable
goal when we consider that a \hbox{1 \mearth}\ planet orbiting in the {\em
habitable zone} of 0.1 \msun\ star induces a $\sim 1.7$ \ms\ radial velocity
amplitude. \cite{riveira05gj876} have nevertheless already estimated
\mbox{$\la1.5$} \ms\ radial velocity jitter for the M4 dwarf, GJ 876 (at optical
wavelengths).

A major cause of radial velocity jitter is the presence of magnetic activity
induced starspots that distort absorption line profiles \citep{saar97spots}. Any
time variant asymmetries in line profiles due to starspots will lead to biased
measurements of dynamically induced radial velocities from orbiting bodies such
as planets. If the radial velocity signal of a planet is of similar magnitude
or smaller than spot induced jitter, any candidate planet signal may not be
recovered with reasonable allocations of observing time. In this paper, we
simulate semi-realistic starspot patterns for M dwarfs. By
making use of starspot size distribution models for the Sun and by considering
indirectly observed starspot distribution patterns for M dwarf stars, we model
the subsequent line profiles from which radial velocity jitter is determined via
cross-correlation. This approach extends similar work carried out by
\cite{desort07rvs} and \cite{reiners10rvs} to more
realistic spot distributions that are solar spot distributions extrapolated to
active stars. The method is more analogous to
the recent work by \cite{lagrange10spots} that investigates
the effect that solar sunspot activity would have on detection thresholds of an
earth-like planet orbiting in the habitable zone. Here, we use our models to
determine the detection thresholds for earth mass habitable zone planets
orbiting M dwarf stars that exhibit different equatorial rotation velocities and
starspot activity levels. In \S \ref{section:modelling_detecting} we discuss
evidence for M dwarf starspot
distributions and introduce our extrapolated solar model. In \S
\ref{section:synthetic_data}, we briefly describe the methods used to
generate line profiles. A comparison of the relative spot amplitudes for
different { photosphere-to-spot temperature contrast ratios (\tps)} is made
in \S \ref{section:optical_ir}. The
radial velocities induced by stars that exhibit the spot distributions
introduced in \S \ref{section:modellingspots} are then investigated in \S
\ref{section:radvels}. Here, we investigate the effects of spots for three
stellar masses at extremes of the expected \tps\ values. We then carry out
detection threshold simulations for Earth-mass planets in \S
\ref{section:detection} before finally considering possibilities that
generalise our specific model cases further in \S \ref{section:discussion}.

\section{M dwarf spot modelling}
\protect\label{section:modelling_detecting}

\subsection{Observed starspot distributions}
\protect\label{section:observedspots}
Compared with earlier spectral types, { our knowledge of starspot distribution
patterns on M dwarfs is less complete.} For earlier
spectral types, publications that predict and report observations of these
distributions are numerous. The solar analogue is our chief reference point for
which we observe spots appearing chiefly at 0\degs\,-\,40\degs\ latitudes. There
is also considerable evidence that other stars exhibit similar starspot
activity to the Sun. Observations made by
the Mount Wilson Survey \citep{baliunas95}, which has observed changes in the
rotation periods of a number of stars { over} decades \citep{donahue96},
suggests
that similar magnetic dynamo processes are at work.
The appearance of spots within defined (low) latitude bands has been
attributed to the interface dynamo process by which the magnetic fields
responsible for the spots are generated. A magnetic dynamo located at the
boundary between the radiative core and
convection zone { together with} radial transport of flux through convection, is able to
explain the appearance of photospheric
flux at low-mid latitudes only \citep{moreno92fluxtubes}. For more rapidly
rotating stars which can be indirectly imaged via the Doppler imaging process,
this scenario \citep{schussler96buoy} is less successful at predicting the
starspot
distributions \citep{barnes98aper}. However spots still often appear within
specific latitude regions in Doppler images.


\begin{figure*}
\begin{center}
\begin{tabular}{ccc}
\includegraphics[width=35mm,angle=270]{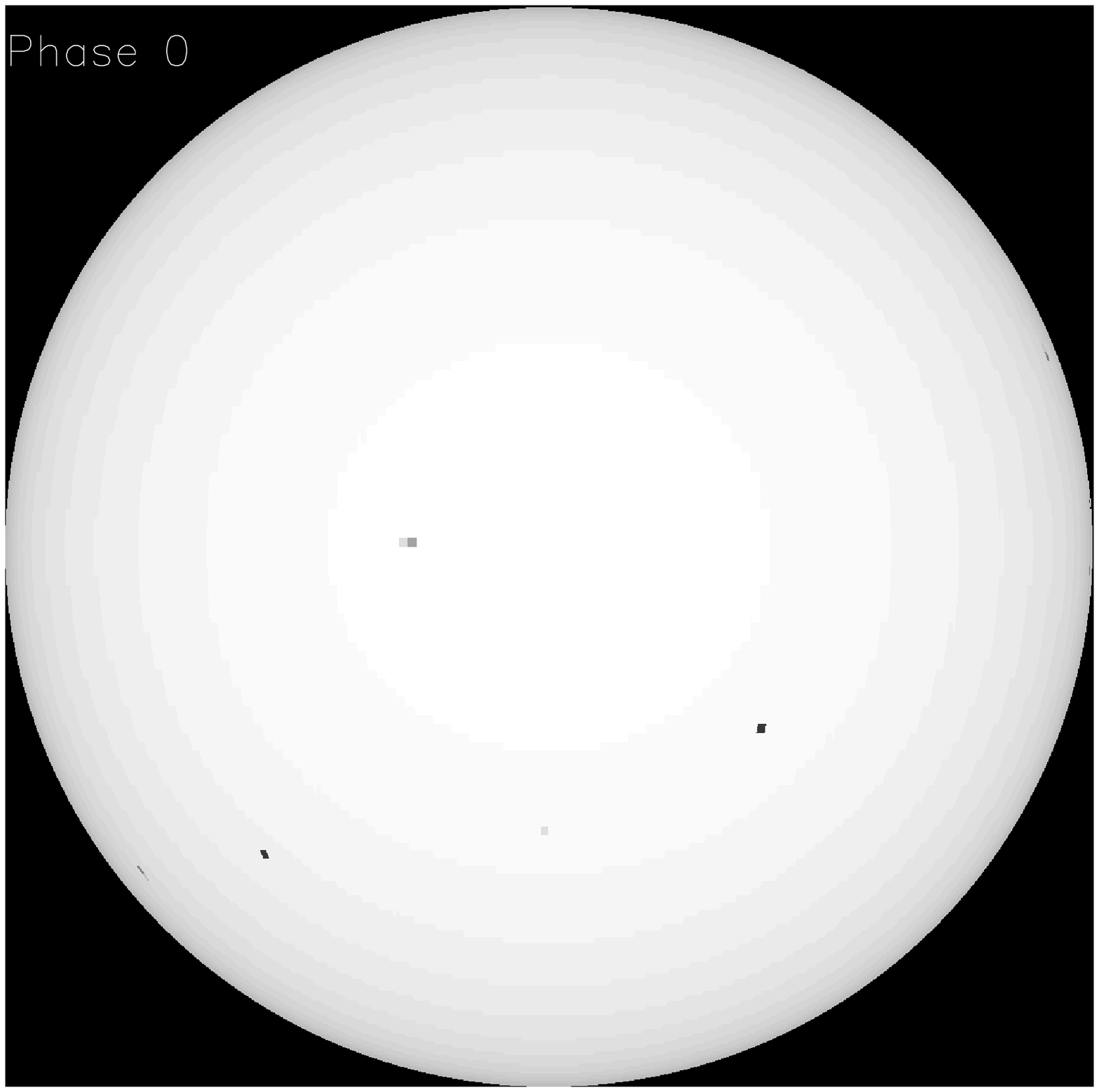}  & \hspace{-3mm}
\includegraphics[width=35mm,angle=270]{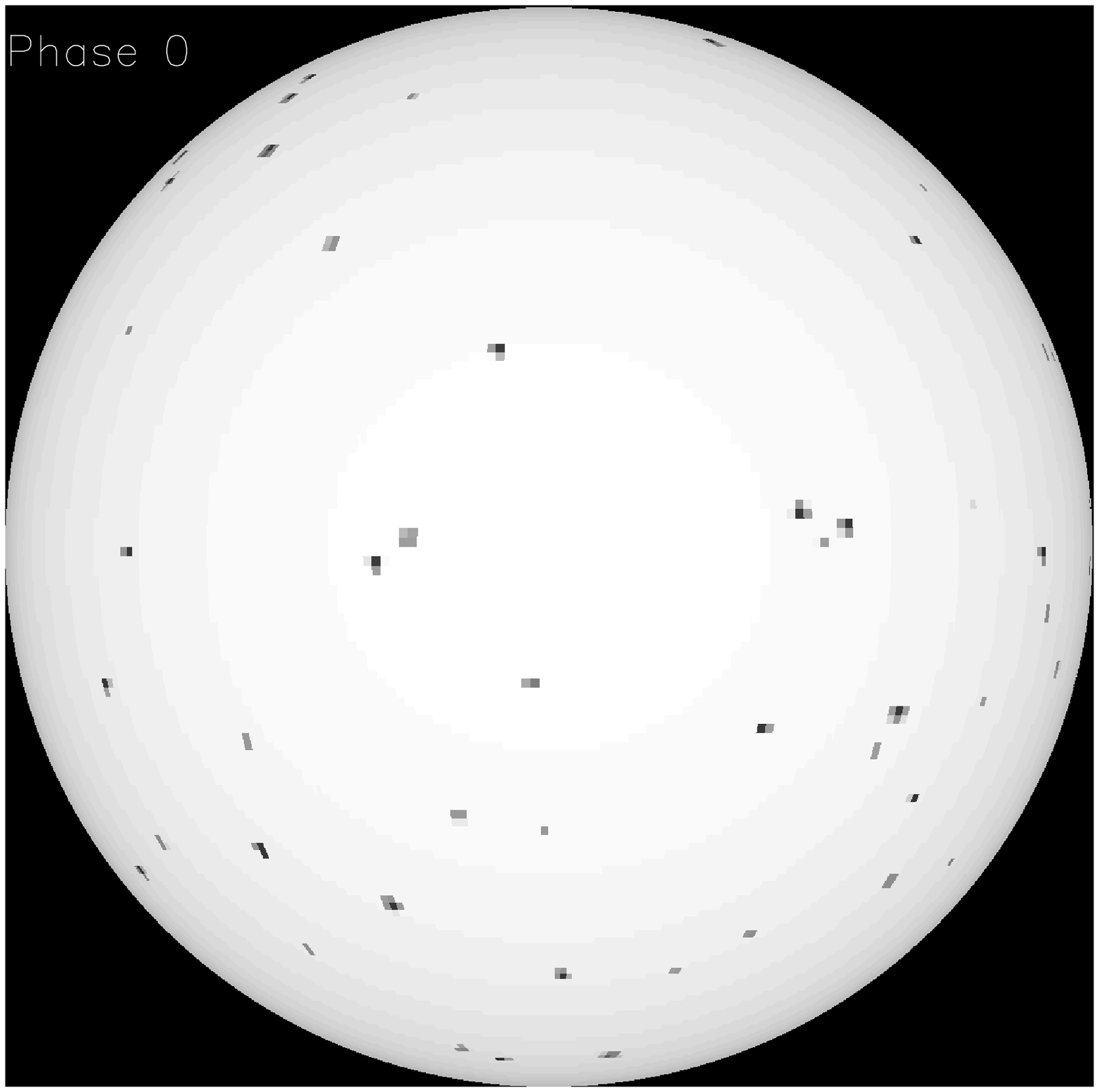}  & \hspace{-3mm}
\includegraphics[width=35mm,angle=270]{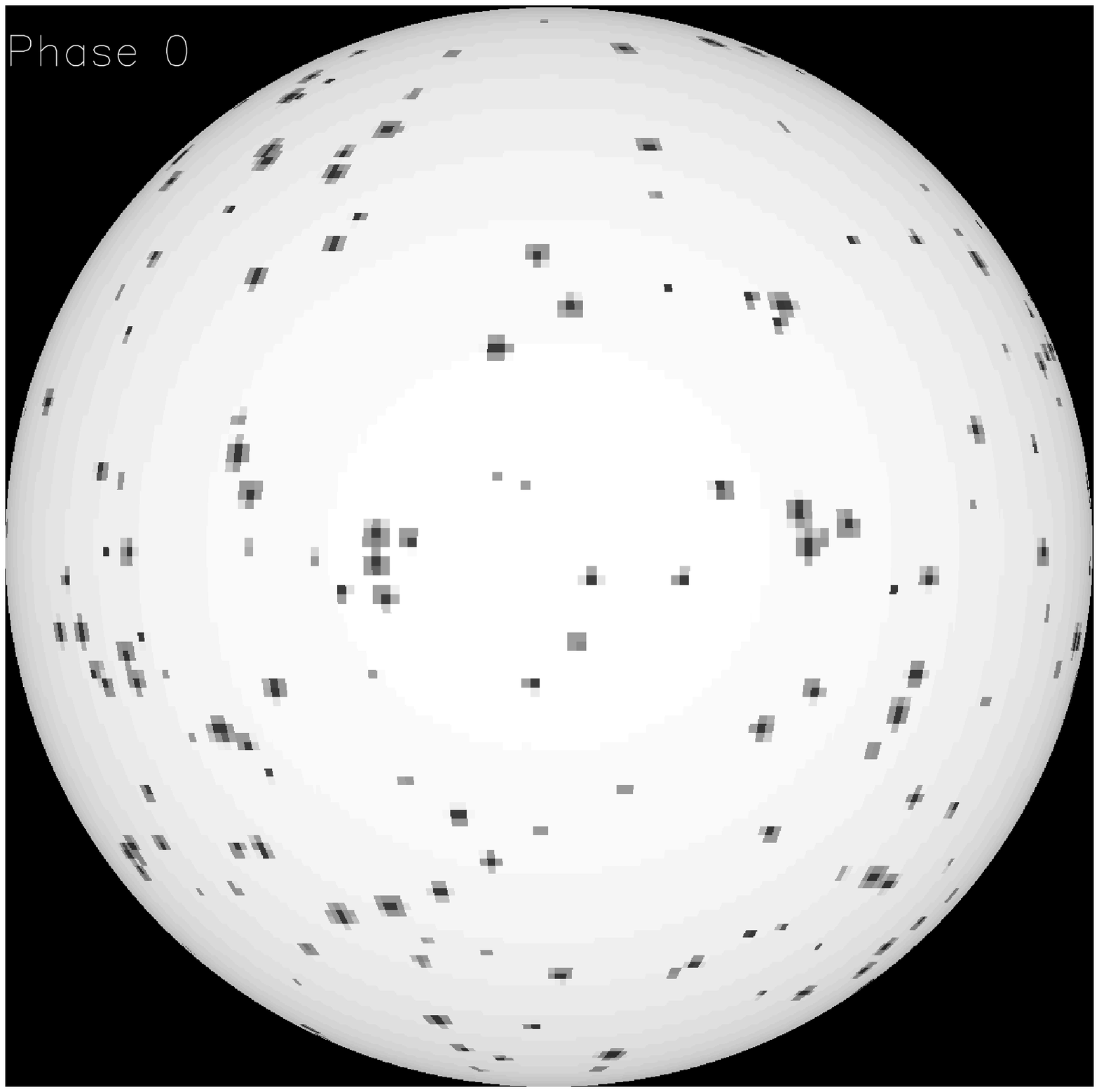}  \\ 
\vspace{-4mm} \\
\includegraphics[width=35mm,angle=270]{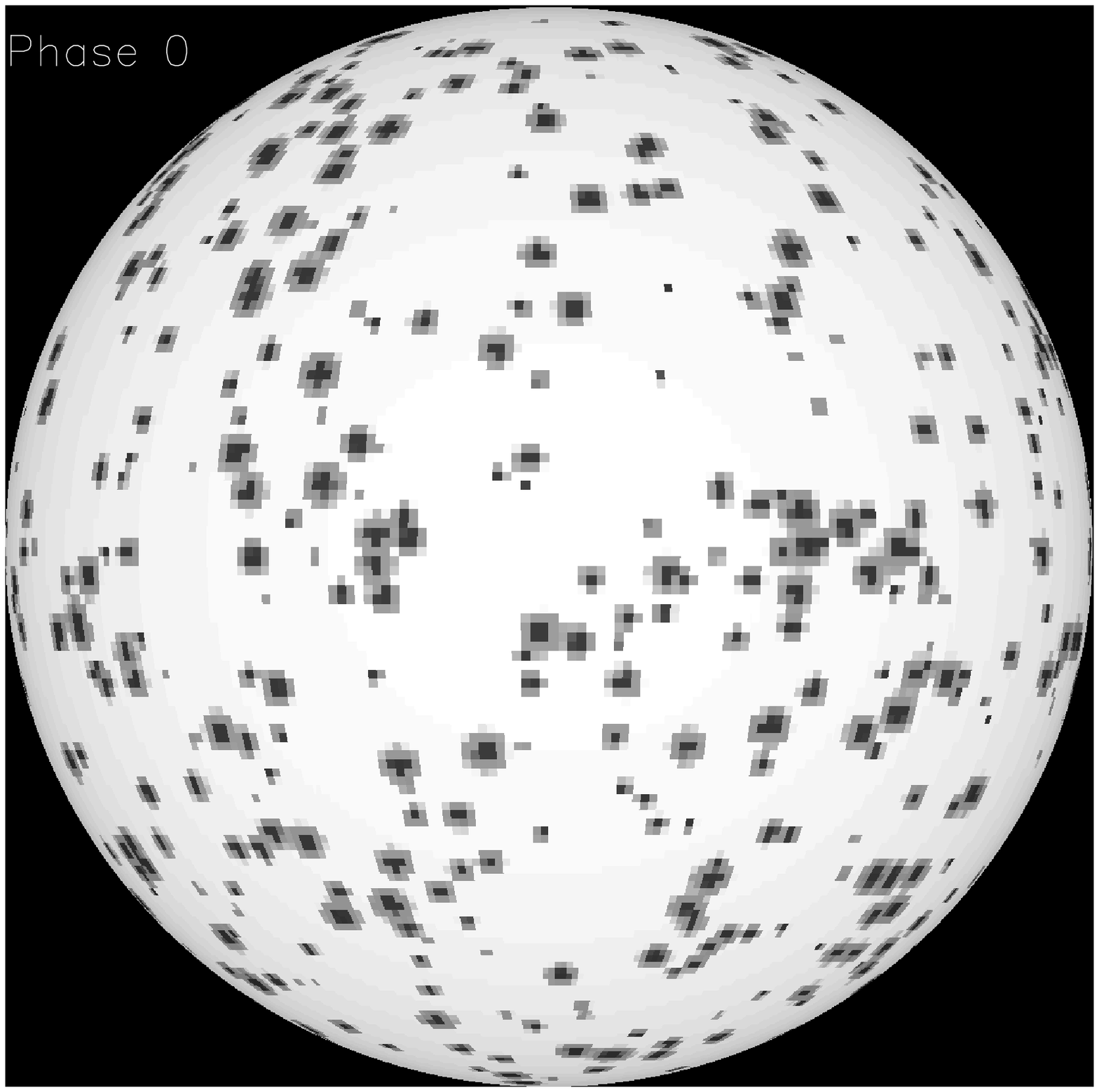}  & \hspace{-3mm}
\includegraphics[width=35mm,angle=270]{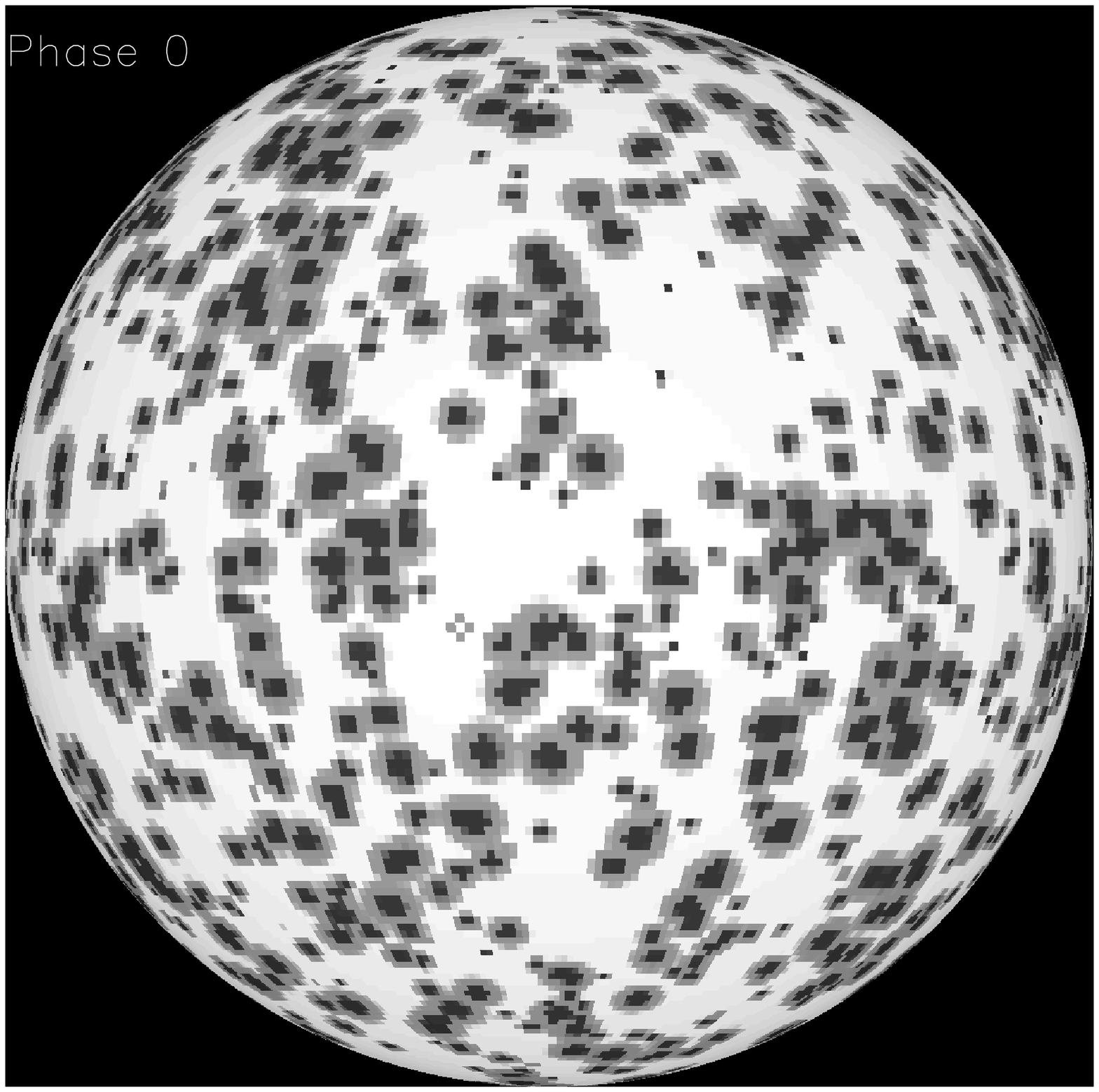}  & \hspace{-3mm}
\includegraphics[width=35mm,angle=270]{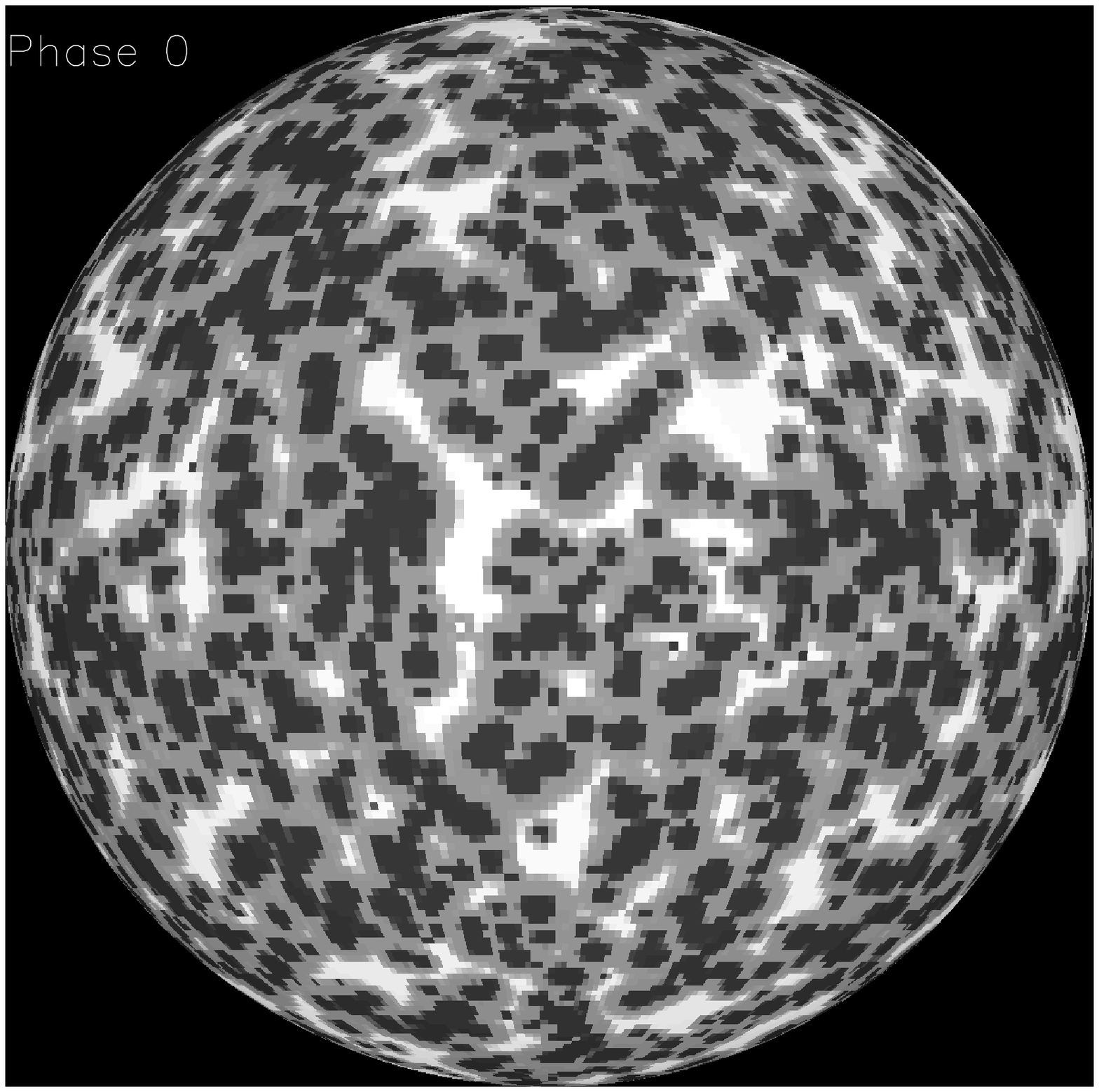}  \\
\end{tabular}
\end{center}   
\vspace{5mm} 
\caption{The distribution of spots for Models 1 (top left) 
to 6 (bottom right) for an M dwarf. Models 1 and 2 are analogous to solar min and
solar max activity levels while Model 3 represents a high solar activity case.
Models 4 to 6 are included for completeness (see Table 1).}
\protect\label{set4t7}
\end{figure*}
\raggedbottom

Unlike G and K stars, owing largely to their inherent faintness, M dwarfs are
less well studied. By mid-M, it is predicted that stars become fully
convective and thus the standard $\alpha\Omega$ interface dynamo process can no
longer operate. There is nevertheless a great deal of evidence indicating
that fully convective M dwarf stars are magnetically active. A study by
\cite{gizis00} found that the
percentage of objects with \ha\ appearing in emission peaks at late spectral
type while \cite{mohanty03activity} found that the equatorial rotation velocity
at which \ha\ activity saturates appears at higher values for mid-late M
spectral types than for early-mid M spectral types. { This was however later
ruled out by \cite{reiners10activity}.} The topic is too large to review at
length here where we are concerned primarily with starspot distributions.

Observations of rapidly rotating early M-dwarfs
via indirect Doppler imaging techniques \citep{barnes01mdwarfs,barnes04hkaqr}
reveal that they are more uniformly covered with spots than earlier spectral
types. \cite{donati08mdwarfs} and \cite{morin08mdwarfs} have found, via
magnetic Doppler imaging using Stokes V measurements, that the magnetic topology
of M dwarfs change at approximately spectral type M4V. Although a switch to a
fully convective dynamo might be expected to lead to a more uniformly spotted
star, the results of \cite{donati08mdwarfs} and \cite{morin08mdwarfs} appear to
counter-intuitively indicate that fields become more dipolar. This phenomenon
has also been
investigated by \cite{reiners09topology} who find that in fact more than 85 per
cent of the magnetic flux is stored in magnetic fields that are invisible to
Stokes V. However, using all Stokes components is necessary for a full
description of a star's magnetic topology. Additionally, the field from the
darkest magnetic regions (e.g. spots) is not visible in Stokes V owing to the
large contrast seen at optical wavelengths. { The high average fields of a
few kG found on M dwarfs { \citep{johnskrull96mdwarfs,reiners07magnetic,reiners10activity} are unlikely}
to be concentrated in small spots because very high local fields would be
necessary. They are perhaps evenly distributed across the surfaces of the stars.
The traditional picture of spots as the only concentrations of magnetic flux may
therefore be misleading in these stars. It is thus clear that} neither magnetic
imaging nor surface brightness imaging at optical wavelengths are able to give a
complete picture of the starspot/magnetic field topology of stars.

While starspot patterns (e.g. the latitudes at which spots appear) on more
rapidly rotating G and K stars may vary as a function of rotation, { it is
not clear whether} such changes take place among {\em fully convective} M
dwarf stars that are expected to generate magnetic fields via a turbulent
dynamo process. Hence those surface brightness images derived for M dwarfs
\citep{barnes01mdwarfs,barnes04hkaqr} may also be representative of slower
rotators. Moreover, at later spectral types, M7\,-\,M9.5 for example,
\cite{reiners10activity} find no correlation between rotation and
magnetic flux generation. Further evidence that starspots may be more uniformly
distributed comes from observations of reduced starspot induced
lightcurve variability. \cite{messina03spots} have shown that already by
spectral type K6\,-\,M4, the maximum starspot induced photometric variability is
around a factor of 2 lower. Other individual studies of mid-M and late-M (M5 \&
M9) dwarfs \citep{rockenfeller06mdwarfs,rockenfeller06m9dwarf} also show
peak-to-peak amplitudes of order 0.05 or less in G, R and I photometric bands. A
more uniform distribution of spots (rather than clustering of spots in one or
two active regions) { would be one explanation for} the observed reduction in
photometric induced light variations in M dwarf stars. 

A series of investigations into spot coverage factors (\citealt{neff95tio,
oneal96tio}, 1998 \& 2004) have been carried out using TiO
as a tracer
of cooler temperatures on G \& K stars of different spectral class, including
dwarfs. These studies, indicate typical spot coverage of 20\,-\,50 per cent for
active stars. This may seem surprising, but indicates that lightcurve analyses
and Doppler imaging studies (that typically find of order 10 per cent spot
coverage) are not sensitive to an underlying spot distribution. It is again
unclear whether these spots arise from a boundary dynamo or whether a turbulent
dynamo is responsible. Based on these findings, investigations of high levels
of spot coverage seem warranted. A threshold level of spot coverage may be
expected (for uniform spot coverage) at which any radial velocity jitter effects
no longer increase in magnitude.

\nocite{oneal98tio,oneal04tio}

With a largely unknown starspot pattern in moderately rotating M dwarfs and the
general prediction that a distributed dynamo should produce randomly distributed
spots, we carry out a number of simulations to assess the detectability of
planets around M stars. { Without strong evidence to the contrary, we assume
that more rapidly rotating stars are more spotted.} Synthetic starspot models
are used to generate line profiles and to investigate the radial velocity
amplitudes resulting from non-uniform line profiles. These radial velocities are
then used to determine our ability to detect low-mass planets which are in
habitable zone orbits around M dwarf stars.

\subsection{Modelling randomly distributed spots}
\protect\label{section:modellingspots}

We synthesize spot maps which follow a log-normal size distribution on the
surface of an immaculate star. The Doppler imaging code ``Doppler
Tomography of Stars'' (DoTS) \citep{cameron97dots} was then used to place the
spots on the surface of a model star. The input parameters to DoTS for modelling
spots were first introduced in \cite{jeffers05activelongs}. They are, where x is
a random number in the range
\hbox{($ 0 \le x \le 1$):}

\medskip
\noindent
(i) { longitude:} randomly distributed between
0$^\circ$ and 360$^\circ$\newline 
(ii) { latitude:} $-\frac{\pi}{2}<
\theta<\frac{\pi}{2}$, following $\theta= \arcsin \left( 2x+1 \right)$ 
with 0$\le$x$\le$1, to eliminate an artificial concentration of spots
at the pole 
\newline 
(iii) { spot radius:} computed using the
previously described log-normal distribution as tabulated in
Table~1 
\newline 
(iv) { spot brightness \& spot
sharpness:} modelled to obtain an umbral to penumbral ratio of 1:3
\citep{solanki99} \\

\begin{table*}   
\protect\label{tab:input}
\begin{tabular}{l c c c c c l }
\hline
\hline
Model & 1 & 2 & 3 & 4 & 5 & \,\,\,6 \\
\hline
{$\sigma_A$} ($\times 10^{-6}A_{1/2 \odot}$) & 3.8 & 5.0 & 6.8 & 9.2 & 12.2 & 15.8 \\
{(dN/dA)$_{max}$} & 5 & 25 & 65 & 125 & 205 & 305 \\
{ Umbral Spot Coverage} & 0.03\% & 0.3\% & 1.6\% & 6.1\% & 18\% & 48\% \\
{ Total} Spot filling & 0.03\% &  0.3\% & 1.9\% & 9.0\% & 29.5\% & 62.4\% \\
\hline 
{Solar:}&{min}&{max}\\
{$R'_{HK}$:}&-5.0& -4.85 & & & \multicolumn{2}{c}{--$>$ TiO obsns.} \\ 
\hline
\hline 
\end{tabular}
\caption{Tabulation of the input parameters to the log-normal size
distribution of star spots given by equation~\ref{equ1}. The parameters
derived by \citet{bogdan88} for the Sun are data sets 1 \& 2, and those
calculated by \citet{solanki99} for active stars are data sets 3-6 
{ ($A_{1/2 \odot} \equiv 2 \pi R^2_\odot$). The spot filling is determined
directly from DoTS and differs from the umbral spot coverage since we also
include penumbral regions. The spot coverage and filling are the same for models 
1 \& 2 since the spot sizes are close to the pixel resolution in these cases.} 
$R'_{HK}$ values corresponding 
to the solar min and max cases are given
\citep{lagrange10spots} and the upper limit for starspot coverage derived
from TiO studies (e.g. \citealt{oneal98tio}) is indicated.}
\end{table*}   
\raggedbottom

The key difference when compared with the solar case is that spots are allowed
at all latitudes. The solar spot size distribution has been determined
by \citet{bogdan88} from direct observations taken from the Mount
Wilson white-light plate collection covering the period 1917-1982.
\citet{bogdan88} show that the number of sunspots, N, { with umbral area,
A, is given by}

\begin{equation}  
\frac{dN}{dA} = \left(\frac{dN}{dA}\right)_{max}
\exp\left(-\frac{(\ln{A}-\ln{\langle A \rangle})^2}{2\ln{\sigma}_A}\right)
\protect\label{equ1}
\end{equation}

\noindent where the constants $\langle$A$\rangle$ and $\sigma_A$ are the mean
and geometric standard deviation of the log-normal distribution { of the sunspot
areas}, and
$\left(\frac{dN}{dA}\right)_{max}$ { is} the maximum value reached by the
distribution.  For the case of the Sun, these values are tabulated in
Table~1 where model 1 is for an inactive Sun and model 2 is for
an active Sun.  Cool, young potentially planet hosting stars are expected to
be more active than the Sun. For these stars we use the
extrapolations of \citet{solanki99} also shown in Table~1. { All starspots
are modelled, following \citet{solanki99} with circular umbral areas. A
starspot is defined such that the umbral region is at the temperature of the
spot (tabulated in Table 2). Following the assumptions of
\citet{solanki99}, we also include penumbral regions with umbral to
penumbral areas of 1:3 (i.e. radii ratios of 1:2). The penumbral regions are
defined such that their intensity is equal to half the difference of the
photosphere and spot intensities.}

We have included the solar values for $R'_{HK}$ at solar min and solar max, as
derived by \cite{lagrange10spots} who used the relationships of \cite{noyes84}
and \cite{lockwood07}. These values also give an indication of the kinds of
activity levels that are routinely found in optical planet survey targets
\citep{tinney02cahk}. { We also indicate the models (i.e. numbers 5-6) that
correspond to the upper limit of $\sim$50 per cent spot filling
fraction determined in active G and K
stars \citep{oneal98tio, oneal04tio}. Consideration of such heavily spotted
stars is of interest, especially if the contrast ratio, \tps\ is also much lower
in mid-late M stars (see \S \ref{section:observedspots} \& \S
\ref{section:radvels}). This may enable more active
stars to be included in radial velocity surveys.}

\section{Generating synthetic radial velocity data}
\protect\label{section:synthetic_data}
The imaging code, DoTS, enables us to generate line profiles using a 3D stellar
model with arbitrarily placed spots. Radial velocity measurements are then made
directly from the line profiles. { DoTS works with two intensity profiles for
each image pixel, representing the photospheric and spot temperatures. The
degree of spot filling for each pixel is represented by a value in the
range 0.0\,-\,1.0, where 1.0 represents complete filling (i.e. an image pixel at
the spot temperature).}

Relative fluxes for a star with a given temperature are determined from the
absolute magnitude and radii determined by \cite{baraffe98} and
\cite{chabrier00} for low
mass stars and substellar objects. Hence we are able to determine the fluxes for
a star of any mass with a specified model photospheric temperature and an
attributed spot temperature. We use two specific photosphere/spot temperature
contrast extremes in later sections of the paper. However, in the following
section we begin by examining different photosphere/spot contrast ratios.
Throughout this paper, we use non-linear limb-darkening values from
\cite{claret00ldc4} to represent the radial variation of intensity for both
photospheric and spot temperatures.

{ 
\subsection{Choice of near infrared passband}

Our choice of infrared passband for the simulations was determined by the
observation that some near infrared wavelength regions contain a greater degree
of spectral line information, in the sense that more sharp features are found in
these regions. This was noted for example by \cite{reiners10rvs} who found that
the highest radial velocity precision was achieved in the Y band when compared
with the J and H bands. The Y band outperformed the J and H bands in terms of
achievable spectral precision for all M dwarf spectral types (see their Fig.
5). Only for the { early to mid} M spectral types, owing to the much higher spectral
information content, does the V band still outperform the infrared bands despite
much lower flux. As we shall see in the next section, the main advantage of
observing at infrared wavelengths is the reduction in starspot induced
jitter. For the following simulations, we therefore focus on the Y band which
appears to offer the best chances of detecting low amplitude signatures that
may arise from the reflex motion of orbiting planets.
}
\raggedbottom

\subsection{Relative visible and near infrared spot induced amplitudes}
\protect\label{section:optical_ir}

We first investigate the effect of contrast variations on the amplitude of
the radial velocities. Observations made at two different wavelengths, the
V-band and Y-band regions (centered at 5450 \AA\ and 10350 \AA\ respectively),
are simulated. We begin by placing a single spot with radius 10\degs\ on the
equator of a star inclined at 90\degs. The stellar line profile is calculated
at a number of phases for which the starspot is visible. { The only differences between 
phases are the intensity (due to foreshortening angle) and radial velocity of the spot contribution to the
integrated line profile.} { Since we
are interested in the relative amplitude of starspot jitter only in this
section, the simulated \vsini\ is not important (we used \vsini\ = 10 \kms),
providing it is greater than the instrumental resolution.}

\begin{figure}
\begin{center}
\includegraphics[width=59mm,angle=270]{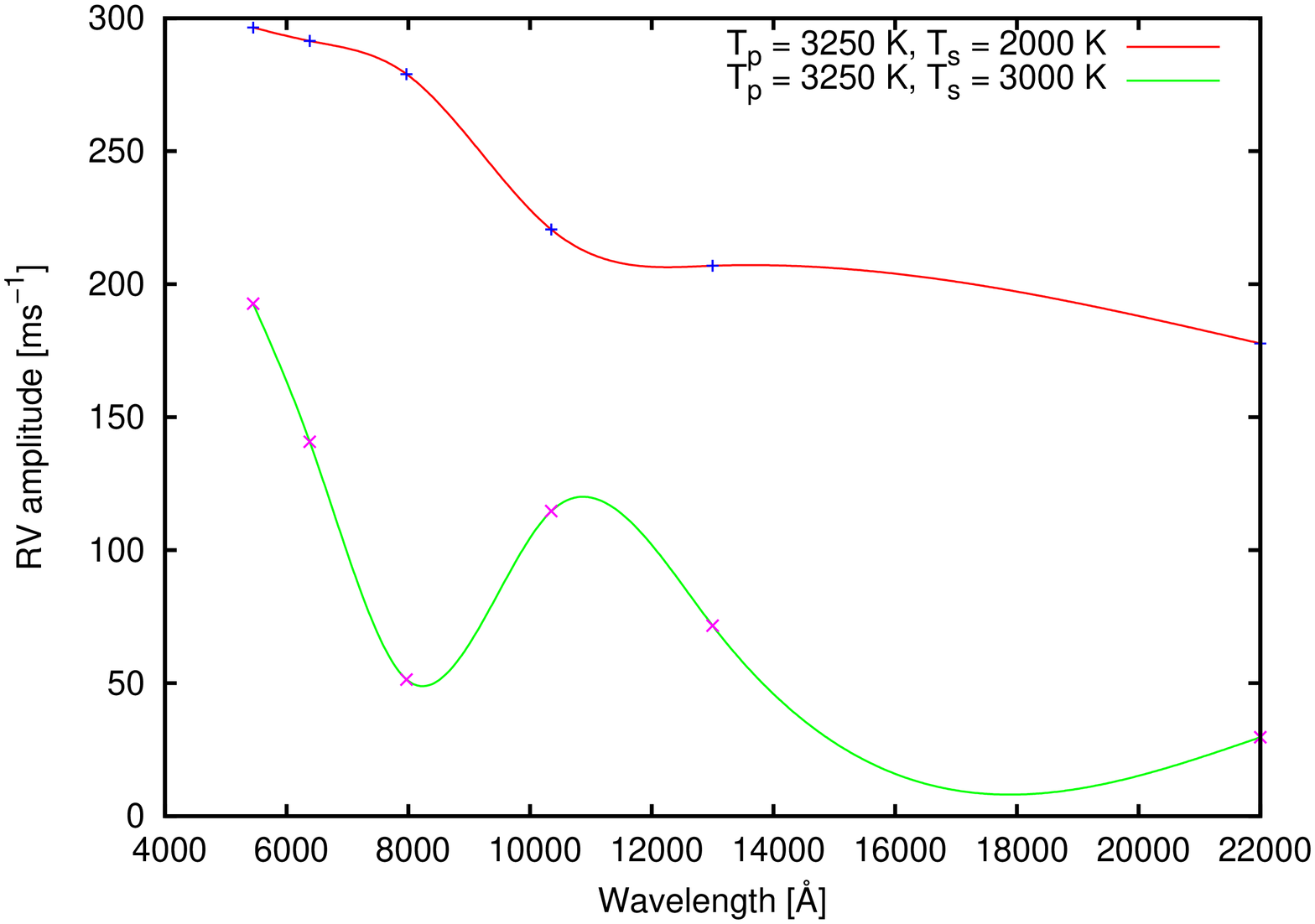} \\
\includegraphics[width=59mm,angle=270]{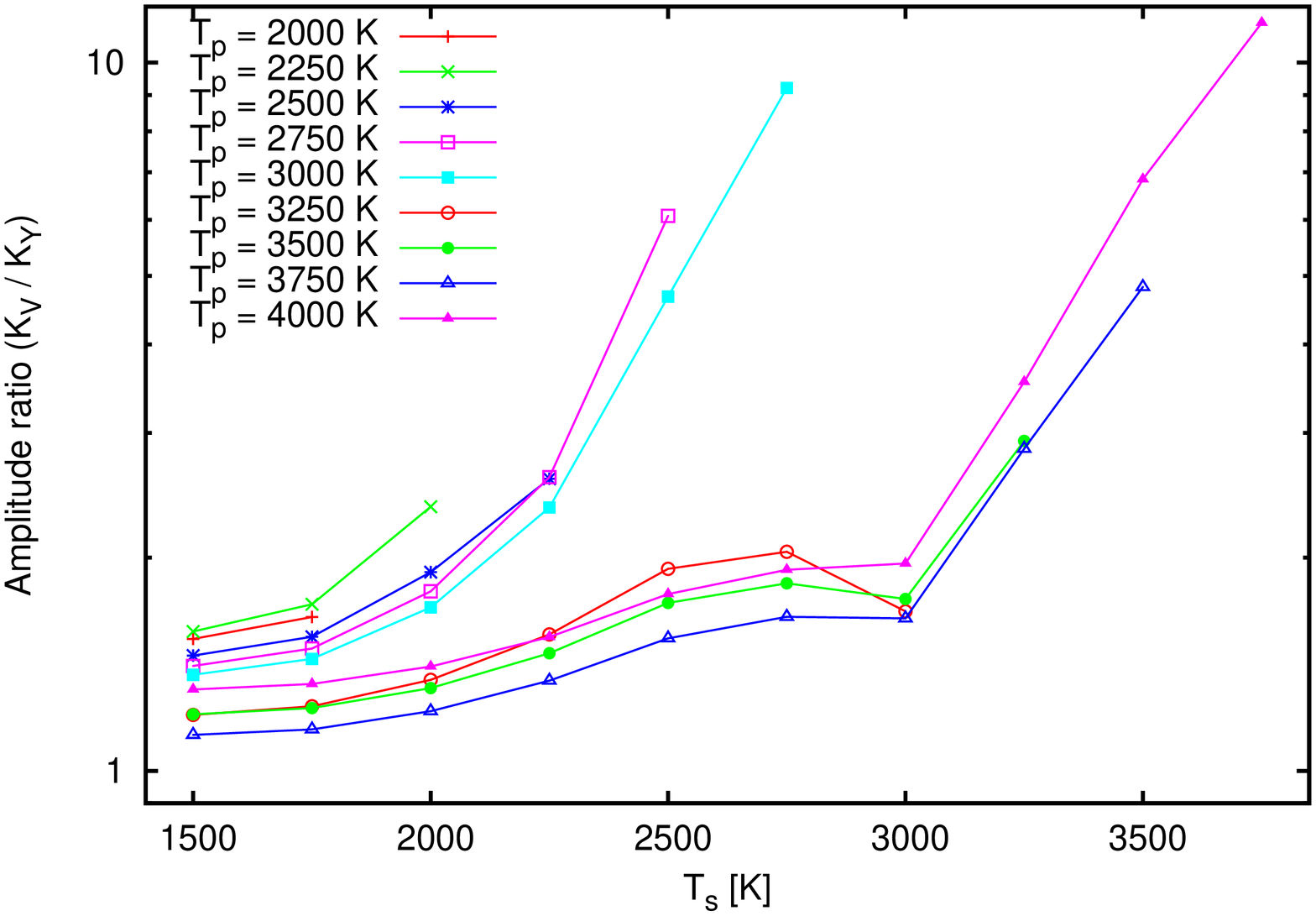} \\
\end{center}   
\vspace{5mm} 

\caption{{ Top: Radial velocity amplitude as a function of wavelength for an equatorial spot of radius 10\degr. The curves are shown for high contrast and low contrast scenarios with $T_p$ = 3250 K. Bottom:} The radial velocity amplitude induced by spots in the visible (V band)
compared with the infrared (Y band). For the labelled photospheric temperatures
in the range 2000 K $\leq$ $T_p$ $\leq$ 4000 K, spot amplitudes ratios are
calculated for 1500 K $\leq$ $T_s$ $\leq$ $T_p$\,-\,250 K.}
\protect\label{fig:ratioplot}
\end{figure} 

{ Using the above parameters, we show in Fig. \ref{fig:ratioplot} (top), the amplitude induced as a function of wavelength for the case where $T_p$ = 3250 K. The curves represent high contrast (\hbox{$T_s$ = 2000 K}) and low contrast (\hbox{$T_s$ = 3000 K}) scenarios. The decrease in amplitude as a function of increasing wavelength is more pronounced for the low contrast case. Here, a secondary effect, namely the relative equivalent width of the lines at the photospheric and spot temperatures, also becomes important and is responsible for the increase in amplitude at \hbox{$\lambda = 10350$ \AA}\ and \hbox{$\lambda = 13000$ \AA}\ when compared with shorter and longer wavelengths. This effect is discussed further below. The results are in broad agreement with those presented in \cite{reiners10rvs}.}

In Fig. \ref{fig:ratioplot} (bottom) we show the relative V-band/Y-band spot induced
amplitudes, {$K_V/K_Y$}, for various spot contrast ratios. {
There are a number of factors that characterise the magnitude of the
starspot radial velocity induced jitter {\em in a given passband, or specific
wavelength range}, the most important being (1) the contrast ratio
between starspot and photosphere and (2) the normalised equivalent width of the
local intensity profile. Additional, secondary effects come from changes in limb
darkening coefficient which we also model using non linear limb darkening values
(see Fig. 3 of \cite{claret00ldc4}). The local intensity profile may be
considered as the local un-broadened  spectrum
of a non-rotating star. The combination of the Doppler shifted local intensity
profiles from all points on the star, or in our case, all pixels on the model
star, are summed up to give the observed profile (it is this summation of
local intensity profiles that is also employed to derive indirectly resolved
Doppler images \citep{vogt83spot,vogt87mem,cameron01mapping} of spotted stars).
The effects of
(1) and (2) are investigated and discussed in detail in \cite{reiners10rvs}.
Here, we summarise those effects in order to illustrate the trends seen in Fig.
2. The Doppler shifted local intensity corresponding to the location of a cool
spot on a star is much less compared with the equivalent photospheric
contribution. The corresponding photosphere/spot intensity
ratio is typically greater than an order of magnitude for standard solar-like
spots that {exhibit $T_p - T_s \sim $ 1000 K}. In this
instance, because the spot contribution to the Doppler broadened
profile is so much smaller than the photospheric contribution, the spot
{\em profile EW and shape} are relatively insignificant. In other words, spots
contribute little flux, both in the line and the continuum, { when the 
contrast between photosphere and spots is large}. Hence the
missing light at a given Doppler shifted position is relatively independent of
wavelength. This is illustrated in Fig. 2 { (bottom)} by the observation that for any
$T_p$, $K_V/K_Y$ is close to unity for the smallest values of $T_s$. The value
of $K_V/K_Y$ increases as the spot temperature (and hence intensity) increases
(i.e. as $T_p/T_s$ decreases). This trend arises simply because the
contrast ratio for fixed
$T_p/T_s$ is much lower in the infrared than in the optical. The wavelength
dependence of the RV signature is illustrated for blackbody fluxes in
\cite{reiners10rvs}. Here we use fluxes derived from the models of
\cite{baraffe98} and \cite{chabrier00} for 5 G yr stars. The top-right point on
each curve in Fig. 2  { (bottom)} indicates $T_p-T_s$ = 250 K. This small temperature
difference leads to starspot induced jitter that is several times smaller in the
Y band as compared with the V band.

\begin{figure}
\begin{center}
\includegraphics[width=60mm,angle=270]{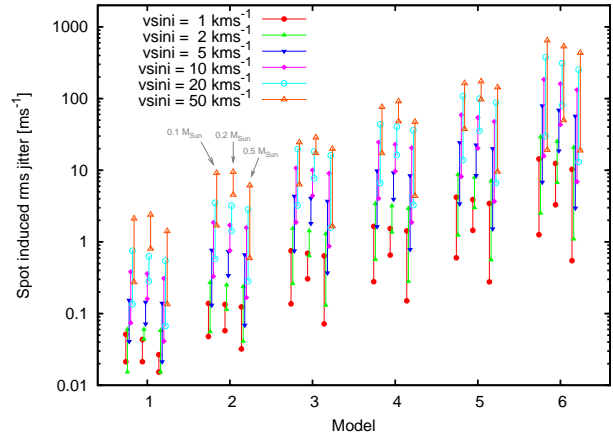}
\end{center}
\caption{{ The { Y band} rms radial velocity variation for each of the 6
simulated spot models. Pairs of values (connected by a line) are shown for each
stellar mass (0.1, 0.2 and 0.5 \msun\ cases are labelled for \hbox{model 2}) and
stellar \vsini~values of 1,\,2,\,5,\,10,\,20 \& 50 \kms. Each pair of values
represents the upper and lower variation for the two extreme $F_p/F_s$ contrasts
listed in Table 2.}}
\protect\label{fig:spot_rms}
\end{figure}
\raggedbottom

For the $T_p$ = 3000 K and $T_p$ = 4000 K cases, reductions in starspot induced
jitter of an order of magnitude are achieved. { These augmented reductions are due
largely to the behaviour of the relative line intensities in the spot and photosphere
at V and Y band wavelengths. While the continuum contrast is low, the relative equivalent 
width and depth of the lines is greater in the V band, compared with the Y band, 
leading to the large observed ratios in Fig. 2.} Again, this
effect is discussed by \cite{reiners10rvs} in \S 4.1.2 of their paper. In
essence, for low contrast ratios, the spot induced jitter is much smaller since
there is little difference between the intensity of the photospheric and spot
contributions. However, the ratio of line depths at $T_p$ and $T_s$ then
becomes important. If the continuum contrast ratio (1) is small but the line
depth of the spectral lines in the spot is {\em greater} than in the
photosphere, the two effects may cancel. In other words, the emission bump
created by the cooler spot spectrum is cancelled by additional absorption due to
deeper lines (see \cite{reiners10rvs}, Fig. 11). { We derive a mean line depth for each
passband region from the models of \cite{brott05}.} For \hbox{$T_p$ = 3250 K}, we see in
Fig. 2 { (bottom)} that $K_V/K_Y$ is much smaller at low $T_p/T_s$. This arises because of a
significant growth in equivalent width of lines in the Y band below this
temperature. For the cases where the continuum contrast ratio is small, the
centre of the lines are then much deeper, leading to higher contrast in the
lines. As we shall see in \S \ref{section:radvels}, this leads to larger
starspot induced velocity amplitudes for our simulated 0.2 \msun\ star in
low contrast scenarios.


Our results in Fig. 2 are in close agreement with those presented in
\cite{reiners10rvs}. A marked difference between the
change in radial velocity induced jitter as a function of wavelength is seen
when moving from their ``toy'' model results of Fig. 11 to the model atmosphere
results (analogous to our results) of Fig. 12. When comparing $K_V/K_Y$,
\cite{reiners10rvs} find an approximate decrease in the starspot induced
jitter of $\sim$2 times \& $\sim$5 times for $T_p = 3700$ K and $T_p = 2800$
K respectively, with $T_p - T_s = 200$ K. We similarly find starspot induced
jitter ratios of $\sim5$ \& $\sim6$ for $T_p = 3750$ K and
\hbox{$T_p = 2750$ K} respectively, with $T_p - T_s = 250$ K. The discrepancies
in the results likely arise from differences in the models, the line depths used
to represent the spot and photosphere local intensities and to a smaller degree
the differences in limb darkening models used. Nevertheless, the results are in
broadly close agreement, indicating a reduction in radial velocity induced
jitter in the Y band over the V band of order $5$ { for $T_p - T_s = 250$ K}.

Fig. \ref{fig:ratioplot} shows a clear advantage of observing at near infrared
wavelengths, when the contrast ratio between spots and photosphere is not
large. At higher contrasts, the advantage is less obvious. Obtaining an estimate
for reasonable values of { $T_s$ vs $T_p$} is not straightforward,
but clearly has important implications for further simulations. This issue is
considered further in the following section.
}

\subsection{Radial velocities induced by random spot distributions}
\protect\label{section:radvels}

The absolute radial velocity amplitudes depend not only on the starspot
distributions but also on the photosphere/spot contrast ratio.
Following \cite{reiners10rvs}, we have investigated two extreme scenarios for
M dwarf stars with three different masses. In Table 2, we list the chief
parameters for the simulated stars, including the photospheric and spot
temperatures.

\begin{table*}
\begin{center}
\begin{tabular}{cccccccccc}
\hline
Stellar Mass	&	$T_{p}$	& $T_{s_1}$	& $T_{s_2}$	&
$P_{Planet}$ & \multicolumn{5}{c}{$K_*$ for planet masses of} \\
		&               &               &               &               
  &1 M$_\oplus$&2 M$_\oplus$&5 M$_\oplus$&10 M$_\oplus$&20 M$_\oplus$\\
M$_{\odot}$	&	[K]	& [K]		& [K]		& [d]	     	   &  \multicolumn{5}{c}{[ms$^{-1}$]} \\
\hline
0.1		&	2750	& 1800		& 2550		& 4.96	     	   & 1.74& 3.48& 8.69& 17.41& 34.76	\\
0.2		&	3250	& 2100		& 3050		& 13.02	     	   & 0.79& 1.59& 3.97& 7.94& 15.87	 \\
0.5		&	3750	& 2450		& 3550		& 35.6	     	   & 0.31& 0.62& 1.54& 3.08& 6.16	\\
\hline
\end{tabular}
\caption{Simulated stellar mass, photospheric temperature, $T_{p}$, spot
temperatures $T_{s_1} = 0.65T_{p}$ { (to the nearest 50 K)} and $T_{s_2} =
T_{p} - 200$ K. Also tabulated are the periods of orbiting habitable zone
planets for each stellar mass and the stellar radial velocity amplitudes induced
by orbiting planets of mass 1, 2, 5, 10 \& 20 M$_\oplus$.}
\end{center}
\protect\label{tab:temps}
\end{table*}

Stars with axial inclination, $i$ = 90\degs\ are simulated according to the
parameters tabulated in Table 2 and for each of the 6 stellar
models described above. A total of 36 line profiles (i.e. every 10\degs) with a
range of \vsini\ values of { 1, 2, 5, 10, 20 \& 50 \kms} are generated for a
complete stellar rotation period. The local intensity profile used to
represent the rotation profile of a non-rotating star is a synthetic Voigt
profile with an instrumental resolution that is appropriate ($R = 70, 000$) for
high resolution IR spectroscopy (e.g. \citealt{jones09rvs_ir}).

{ Figure \ref{fig:spot_rms} plots the starspot induced rms radial velocity
variation for each spot model (introduced in \S 2.2) and \vsini\ value.  The
pairs of values for
each stellar mass of 0.1, 0.2 and 0.5 (points connected by lines in Fig. 3)}
represent the $T_{s_1} = 0.65T_{p}$ (greater rms RV) and $T_{s_2} = T_{p} - 200$
K (lesser rms RV). As expected, there is little variation due to the stellar
mass for the $T_{s_1} = 0.65T_{p}$ point since the relative contrast ratio is
unchanged. The lower point in each case shows more variation for a given model
since the ratio $T_{s_2} = T_{p} - 200$ K is not fixed. { Moreover, for
the $T_p/T_s$ combination of 3250 K / 3050 K, we see that the radial velocity
variation is between 1.5 and 6 times larger than for the corresponding 3750 K /
3550 K case. This difference is due to the effects discussed in the \S 3.2, where the
equivalent width of the lines grows for temperatures close to, but below 3250
K. The large range in factors of between 1.5\,-\,6 is likely due to the spot sizes and number that define each model
and their relationship with the Doppler/spatial resolution limit which improves (i.e. smaller spots resolved) with increasing \vsini.}
This will ensure that for low contrast spots on our 0.2 \msun\ model, the
starspot jitter will be greater than for the 0.1 \msun\ and 0.5 \msun\ models.


{ For model 1, with \vsini\ = 2 kms\ (i.e. a slow rotator analogous to the
Sun) the approximate range of velocities (considering all three stellar masses)
for the two \tps\
extremes are \hbox{$\sim$2\,-\,6 cms$^{-1}$}, while for model 2 the range of
observed values is \hbox{$\sim$ 4 cms$^{-1}$ - 27 ms$^{-1}$}. This is an order
of magnitude smaller than the values observed for the Sun by
\cite{lagrange10spots} which show equivalent radial velocity jitter of solar
minimum and maximum periods of up to 0.6 \ms\ and 2 \ms\ respectively. The
reduction in contrast ratio in the
optical vs the infrared, and the fact that spots may appear at all latitudes,
act to reduce the simulated radial velocity jitter for models 1 and 2. This
latter point is important since placing only a few small spots {\em randomly} at
all latitudes, as opposed to restricting them to solar-like low-latitude bands
may lead to significant reductions in the measured jitter (i.e. spots at low
latitude induce the greatest jitter). By the time a star reaches \vsini\ = 10
\kms, we find rms velocities of \hbox{$\sim$ 4\,-\,38 cms$^{-1}$} and
\hbox{$\sim$ 17 cms$^{-1}$ - 1.9 \ms} for models 1 and 2 respectively.} Clearly,
the solar minimum and solar maximum analogue cases (models 1 and 2) will enable
the greatest possibility of detecting lower-mass planets since they yield the
lowest rms RV variations. As expected, the rms jitter peaks at
Model 6 { which corresponds to a star with a 62.4 per cent starspot filling
fraction. 

For those early M dwarfs that are fast rotators, the spot sizes derived from
Doppler images \citep{barnes01mdwarfs,barnes04hkaqr} are much larger than is
seen for the Sun. It is not clear however whether the spots seen in Doppler
images are actual individual spots or indeed unresolved spot {\em groups}.
Indeed if we take Model 6 and reconstruct the surface spot distribution for a
moderately rotating case, we derive images resembling the Doppler images of
\cite{barnes01mdwarfs,barnes04hkaqr}. 
We have also investigated the effect of larger spots on the radial velocities by
scaling the sizes used for models 1\,-\,6 by factors of 2, 5 \& 10. In other
words, a model identical to each of those shown in Fig. \ref{set4t7} was
created, but with all spots scaled up in radius. The starspot induced rms jitter
maximum in Fig. \ref{fig:spot_rms} is found to increase by a factor of $\sim$ 2
for a spot size scale factor of 10. The main effect however is a change in the
model number at which the peak jitter occurs. For scale factor 2, the rms jitter
is less peaked for model 6 when compared with model 5. For scale factor 5, a
peak occurs between models 3 \& 4, while the peak is at model 3 by scale factor
10. However, even for the more modest scale factor 2 scenario, the planet
detection thresholds (see \S 4) for model 2 will more closely resemble those of
model 3 where the spot scale factor was unaltered. In the following sections,
we do not scale the spot sizes, but use the spot size distributions as
previously defined in \S 2.2
}

\subsection{Stellar mass vs \vsini\ and spot filling}

The effect of \vsini\ is an important factor that must be considered when
determining the RV jitter. While Fig. 3 shows that model 1 does not yield jitter
of $>$ 1 \ms, even at \vsini\ = 20 \kms, a very active star such as that
represented by Model 6 only yields jitter of order 1 \ms\ or less for the low
contrast case when \vsini\ = 1 \kms. There is approximately a 1:1 correspondence
between jitter and \vsini\ for models 3\,-\,6. In other words, doubling the
\vsini\ doubles the rms jitter. However, for models 1\,-\,2, the increment is
small owing to the finite spectral resolution of 70,000 which corresponds to 4.3
\kms.

Since the rotation-activity relation seen at earlier spectral types is
also seen to persist in M dwarfs, through observation of
\ha\ \citep{mohanty03activity}, it seems reasonable to assume that the slower 
rotators will be the least active stars while the more rapid rotators exhibit
higher degrees of spot coverage. Unfortunately, since there have been no
successful studies of spottedness of M dwarfs \citep{oneal05vo}, relating
\vsini\ to the spot filling fraction is not possible. We must therefore make
assumptions about the degree of spottedness as a function of rotation
velocity. Based on solar observations and on the spot filling fraction measured
on other stars \citep{oneal96tio,oneal98tio,oneal04tio}, we assume that
models 1-2 are appropriate for \vsini\ = 2 \kms, and models 3-6 are
appropriate for \vsini\ = 5\,-\,50 \kms. 

\begin{figure}
\begin{center}
\includegraphics[width=59mm,angle=270]{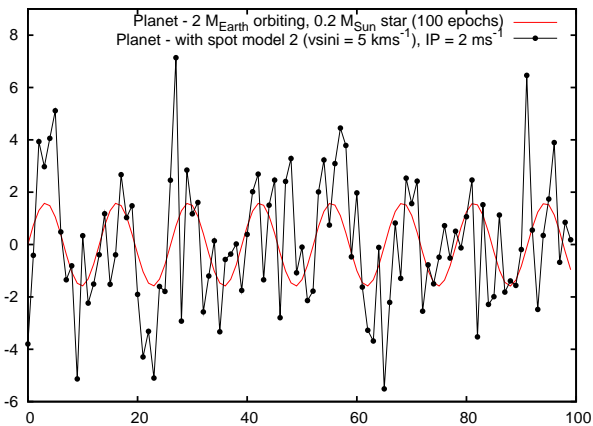} \\
\end{center}   
\vspace{5mm} 

\caption{Example of a simulated stellar RV curve for a 2 \mearth\ planet
orbiting a 0.2 \msun\ star (red/solid line). The RV curve is shown with added
jitter for an instrumental resolution (IP) of 2 \kms, stellar \vsini\ = 5 \kms,
and starspot model 2.}
\protect\label{fig:example_planet}
\end{figure}

{ \cite{jenkins09mdwarfs} have recently determined the \vsini\ values for a
number of M dwarfs, thereby extending the sample of measured rotation values.
This study reinforced the observation
\citep{delfosse98mdwarfs,mohanty03activity,reiners08activity} 
that amongst field
M dwarfs, the earlier spectral types exhibit slower rotation compared with mid
and late M spectral types. In other words, while earlier spectral types are more
likely to be slower rotators, enabling more precise radial velocities to be
determined, the higher mass of the star sets a lower limit on the reflex motion
due to an orbiting planet of given mass. Conversely, the detectable mass limit
for an orbiting planet is adversely affected by the more moderate rotation of
mid to late M dwarf stars, while the lower mass is more favourable for detecting
low mass planets.

For spectral types M2V and M4V (0.5 \msun\ and 0.2 \msun\ respectively),
\cite{jenkins09mdwarfs} (see their Fig. 9) find the median \vsini\ is 3\,-\,5
\kms\ whereas for spectral types M6V (0.1 \msun), \vsini\ $\simeq$8 \kms.
However, velocities from 1 \kms\ to several 10s of \kms\ are found for all stars
in the spectral range considered, albeit with fewer slow rotators found at later
spectral types (i.e. for masses $\leq$ 0.1 \msun\ (M6V)). We therefore
simulate a range of rotation velocities for each spectral type, but urge the
reader to bear in mind the trend of increasing mean \vsini\ with decreasing
stellar mass.

}

\begin{figure*}
\begin{center}
\begin{tabular}{cc}

\includegraphics[bbllx=70,bblly=100,bburx=510,bbury=750,width=54mm,height=85mm,angle=270]{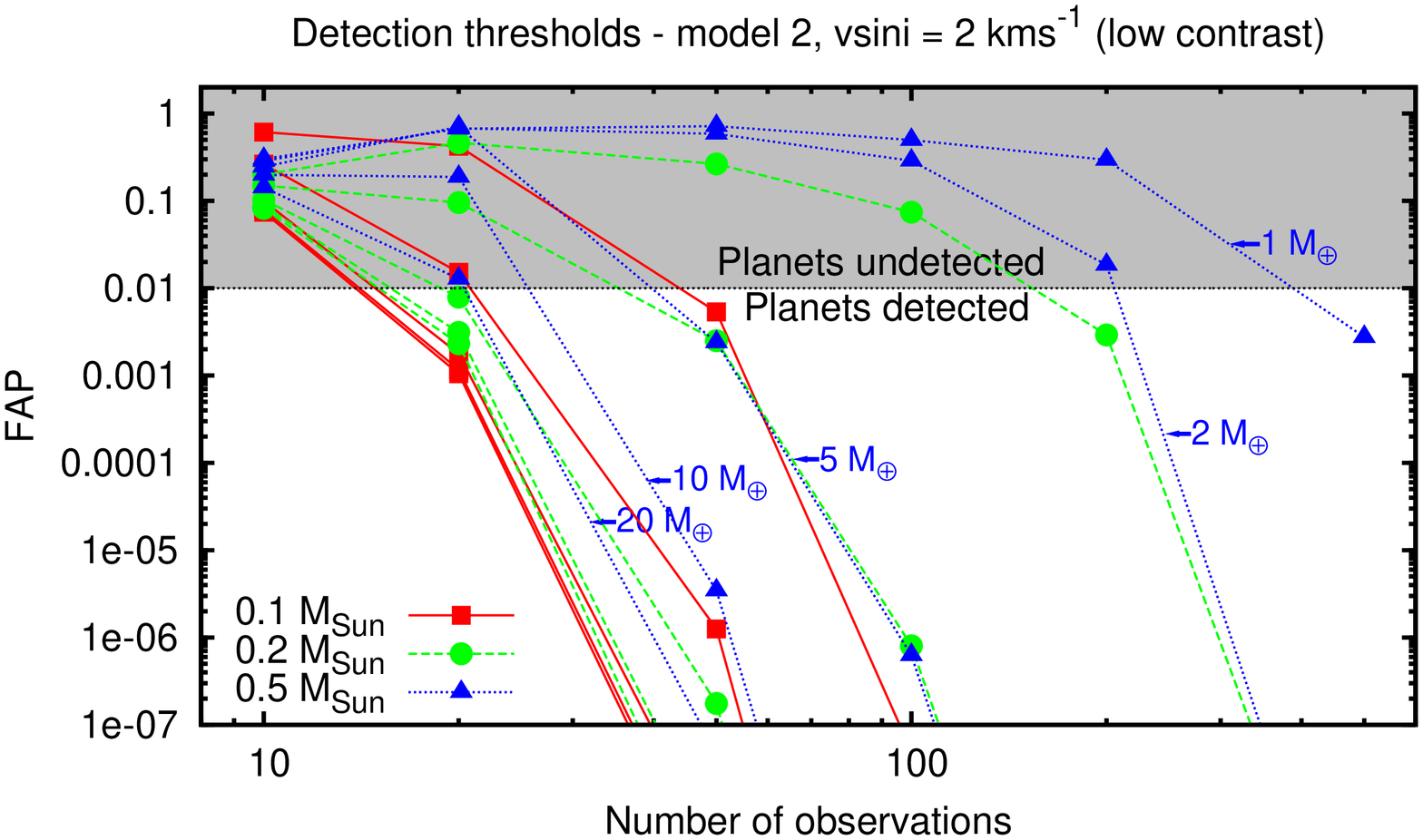}
 & 
\includegraphics[bbllx=70,bblly=100,bburx=510,bbury=750,width=54mm,height=85mm,angle=270]{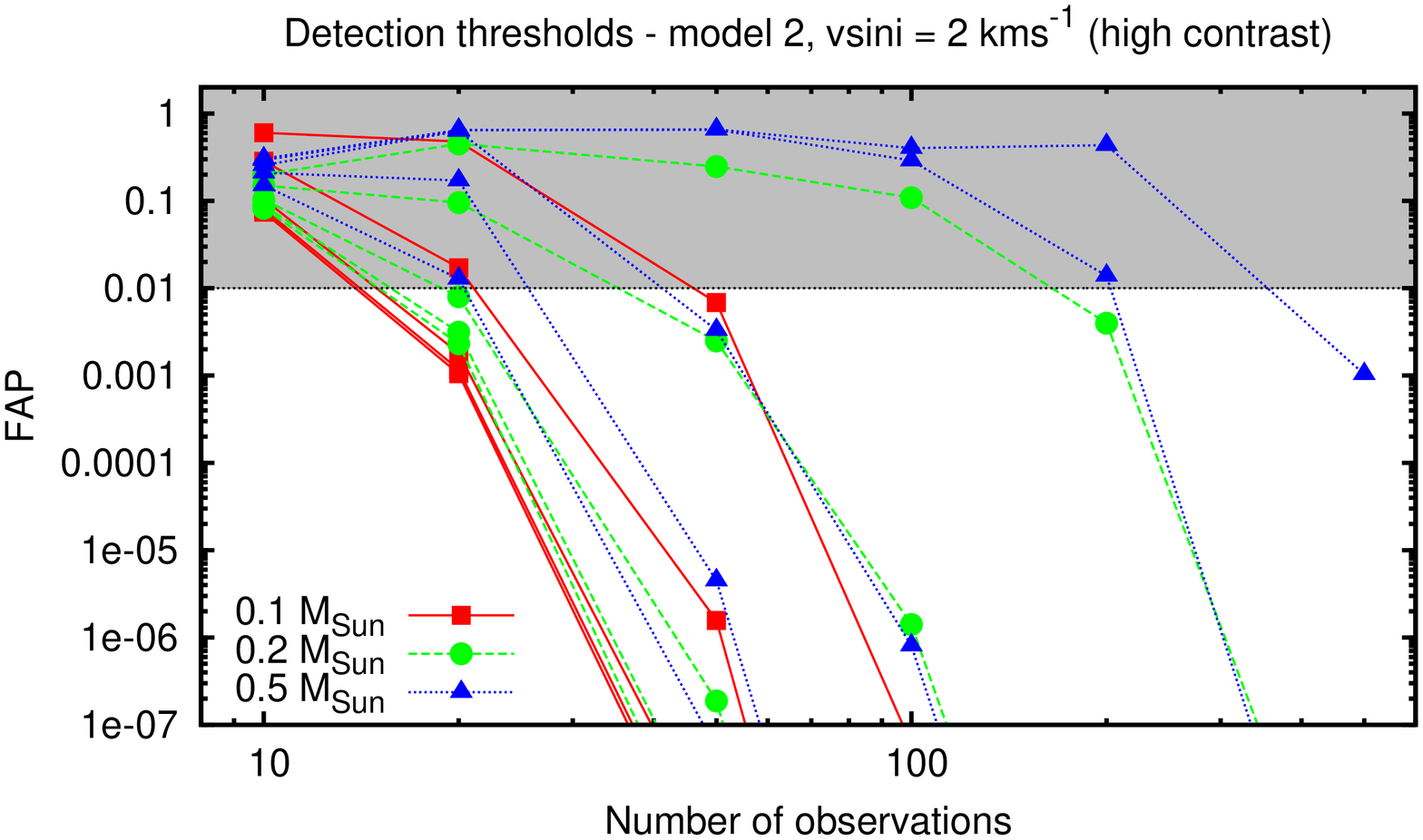} \\

\end{tabular}
\end{center}   
\vspace{5mm} 
\caption{Detection false alarm probabilities (FAPs) vs number of observation
epochs for {\em habitable zone} planets orbiting $M$ = 0.1 M$_{\odot}$
(squares connected by solid/red lines) 0.2 M$_{\odot}$ (circles connected by
dashed/green lines) and 0.5 M$_{\odot}$ (triangles connected by dotted/blue
lines) stars. The five curves for each stellar mass represent planetary masses
of 20, 10, 5, 2 \& 1 \mearth\ decreasing in a left-to-right sense (labelled for
the 0.5 M$_{\odot}$ curves). The horizontal line indicates the 1 per cent false
alarm probability (FAP = 0.01) with the grey region representing undetected
planets. All points with FAP $<$ 0.01 are considered as detections of the
planet. The plots are for activity model 2 with \vsini\ = 2 \kms, instrumental
precision = 1.5 \ms, and with
starspot contrast ratios of $T_s = 0.65T_p$ (high contrast) on the left and $T_s
= T_p - 200$ K (low contrast) on the right.\vspace{-5mm}}
\protect\label{fig:planet_sim1}
\end{figure*} 

\begin{figure*}
\begin{center}
\begin{tabular}{cc}

\includegraphics[bbllx=80,bblly=100,bburx=510,bbury=750,width=54mm,height=85mm,angle=270]{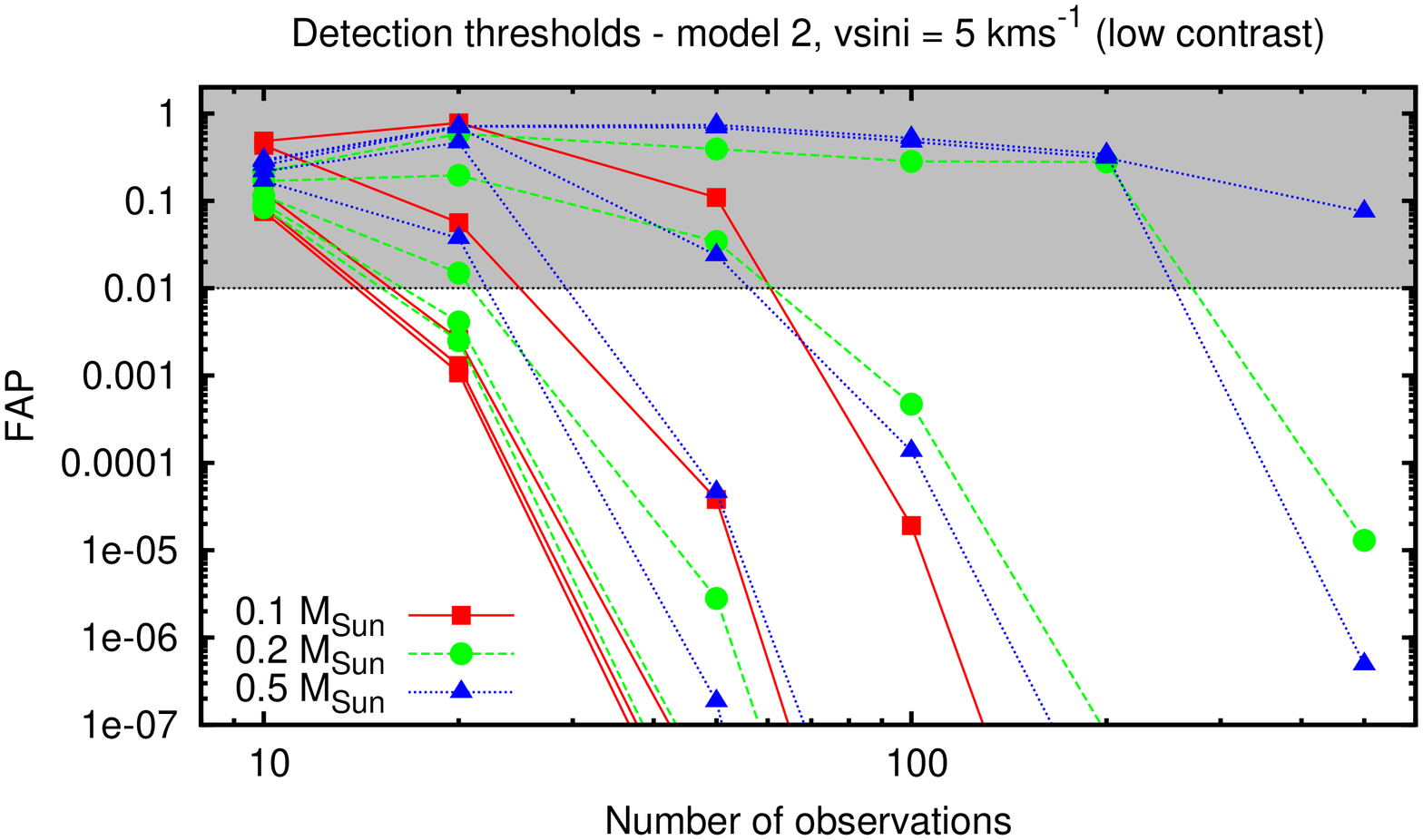}
 & 
\includegraphics[bbllx=70,bblly=100,bburx=510,bbury=750,width=54mm,height=85mm,angle=270]{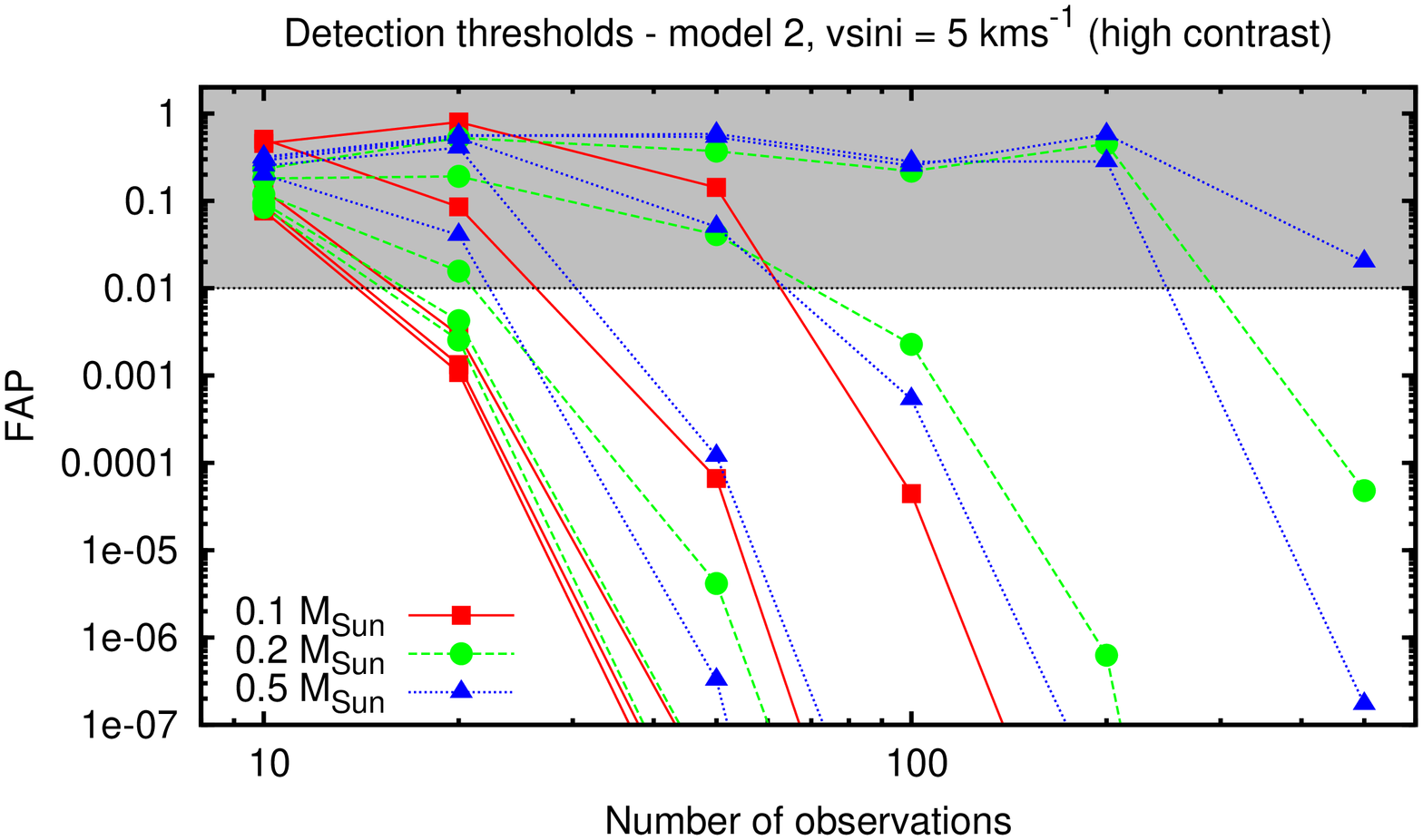} \\

\vspace{-5mm} \\
\includegraphics[bbllx=70,bblly=100,bburx=510,bbury=750,width=54mm,height=85mm,angle=270]{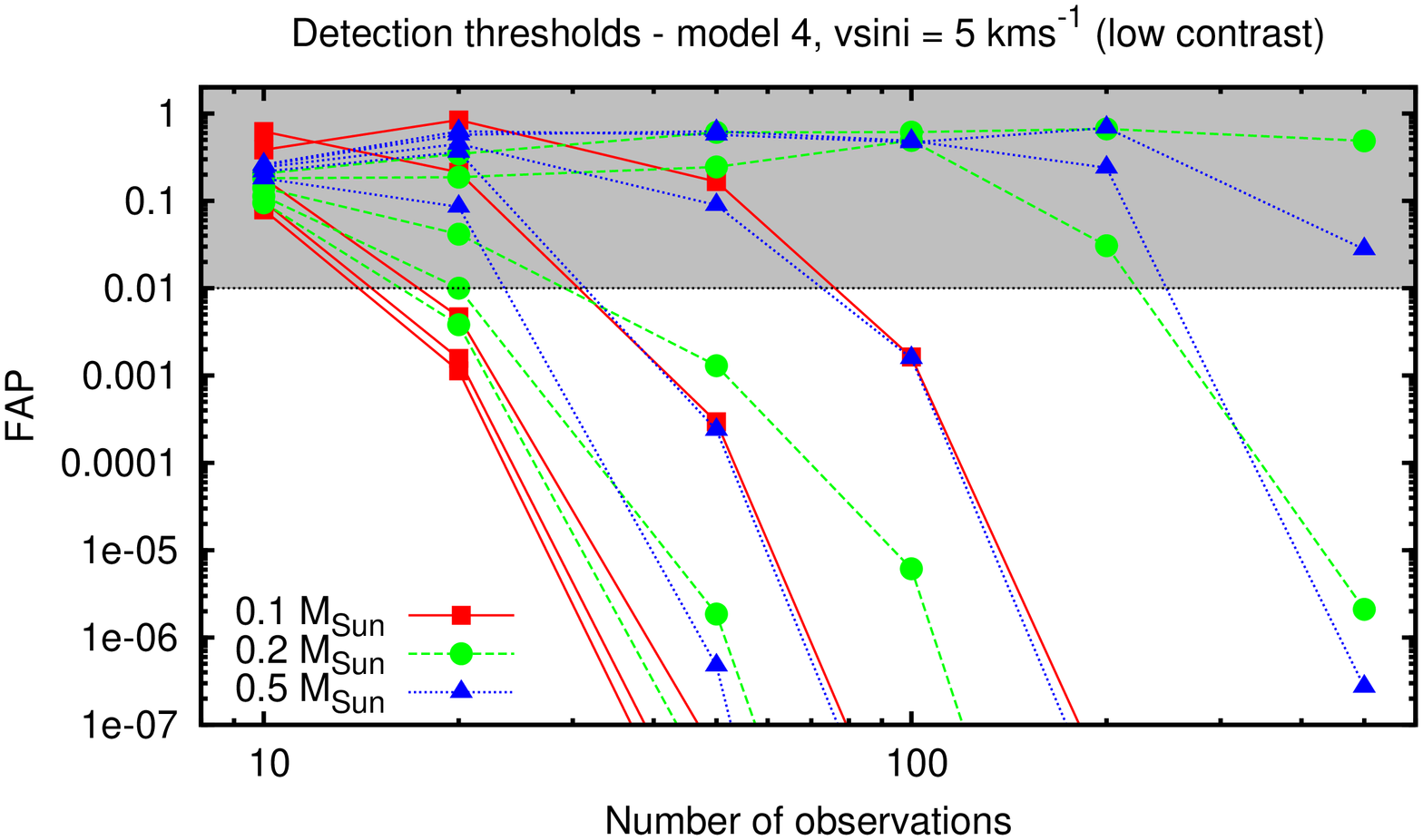}
 & 
\includegraphics[bbllx=70,bblly=100,bburx=510,bbury=750,width=54mm,height=85mm,angle=270]{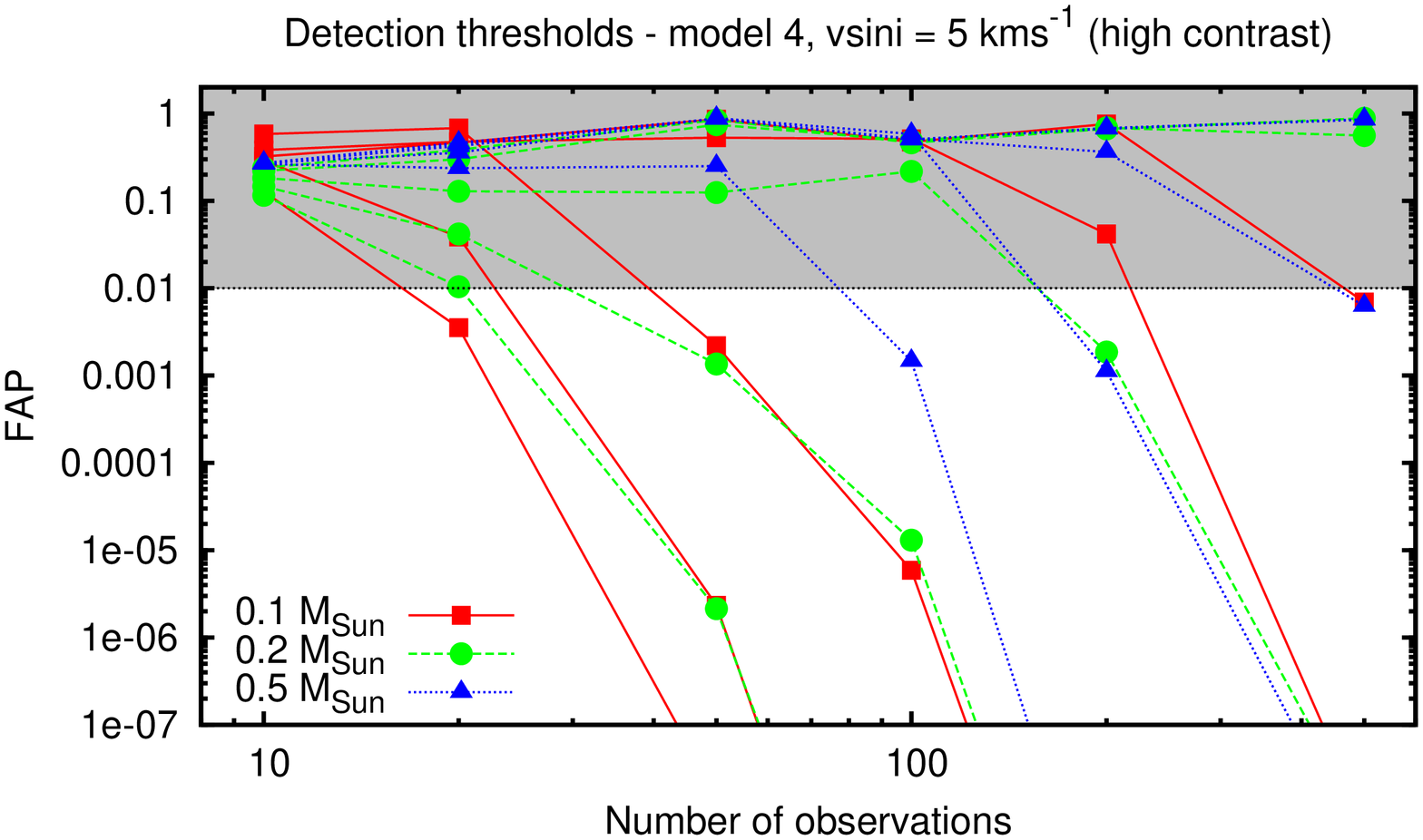} \\

\end{tabular}
\end{center}   
\vspace{2mm} 
\caption{As for Fig. 5 for activity models 2 (top) and 4 (bottom) with
\vsini\ = 5 \kms (instrumental precision = 2 \kms).}
\protect\label{fig:planet_sim2}
\end{figure*}

\begin{figure*}
\begin{center}
\begin{tabular}{cc}

\includegraphics[bbllx=80,bblly=100,bburx=510,bbury=750,width=54mm,height=85mm,angle=270]{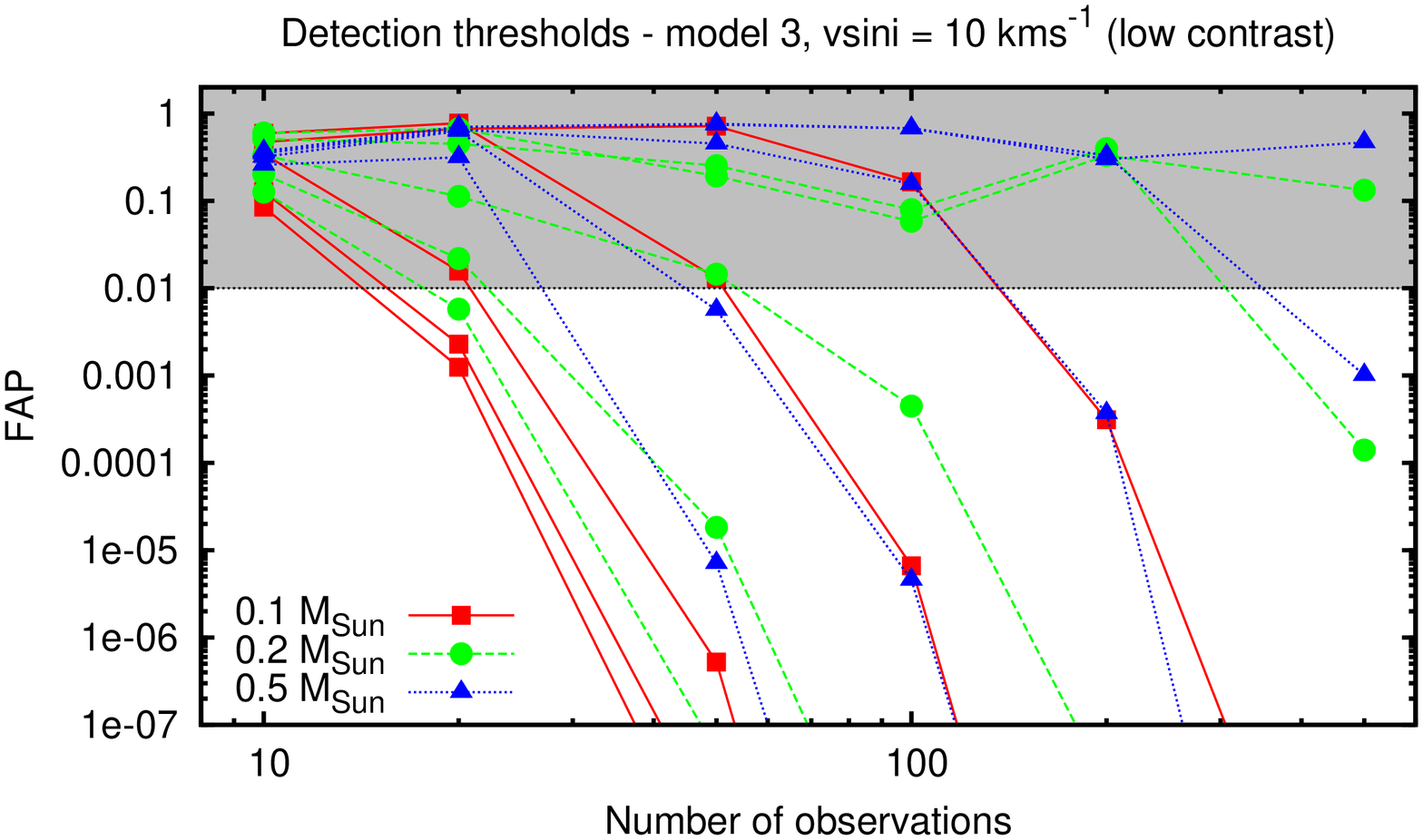}
 & 
\includegraphics[bbllx=80,bblly=100,bburx=510,bbury=750,width=54mm,height=85mm,angle=270]{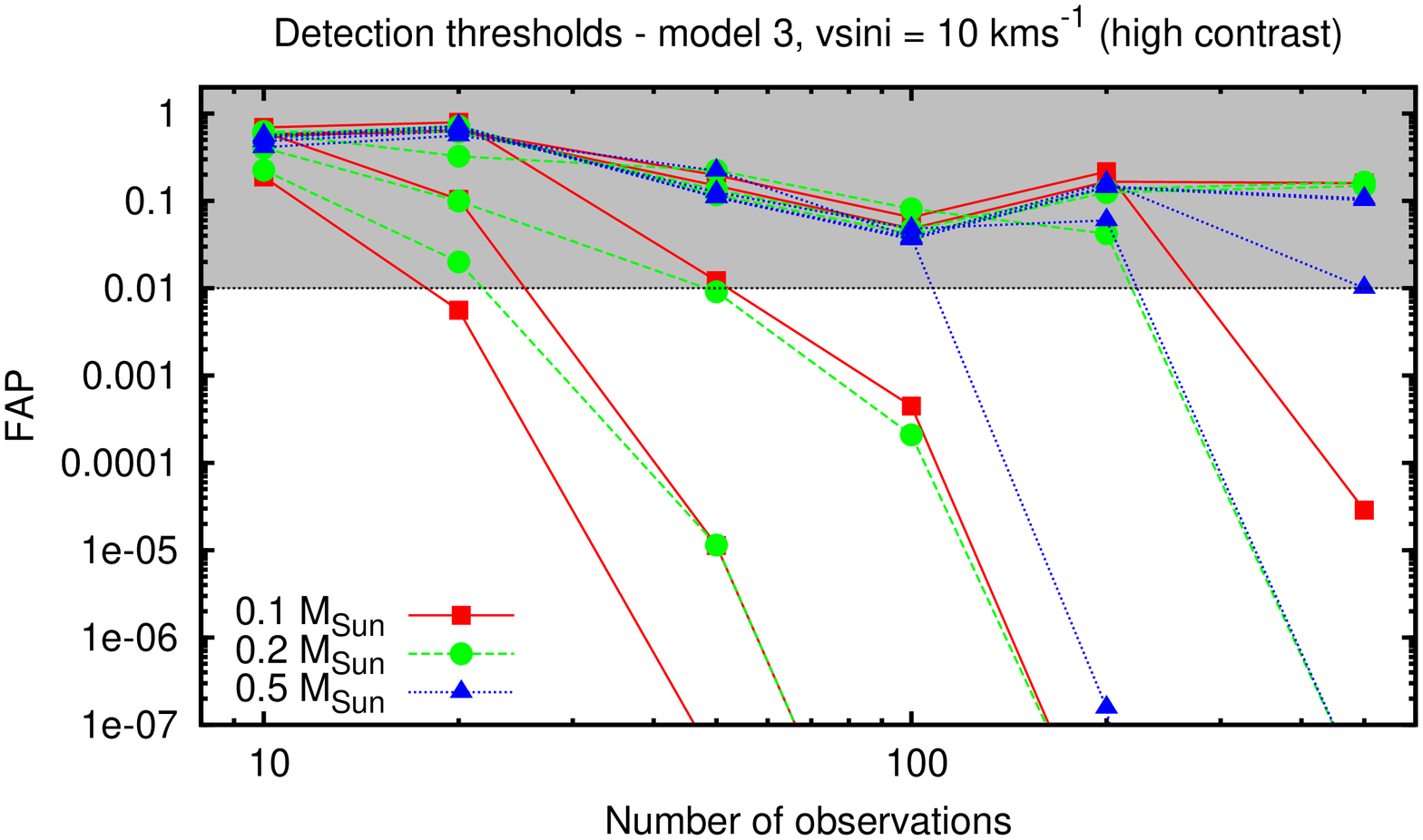} \\
\vspace{-6mm} \\
\includegraphics[bbllx=80,bblly=100,bburx=510,bbury=750,width=54mm,height=85mm,angle=270]{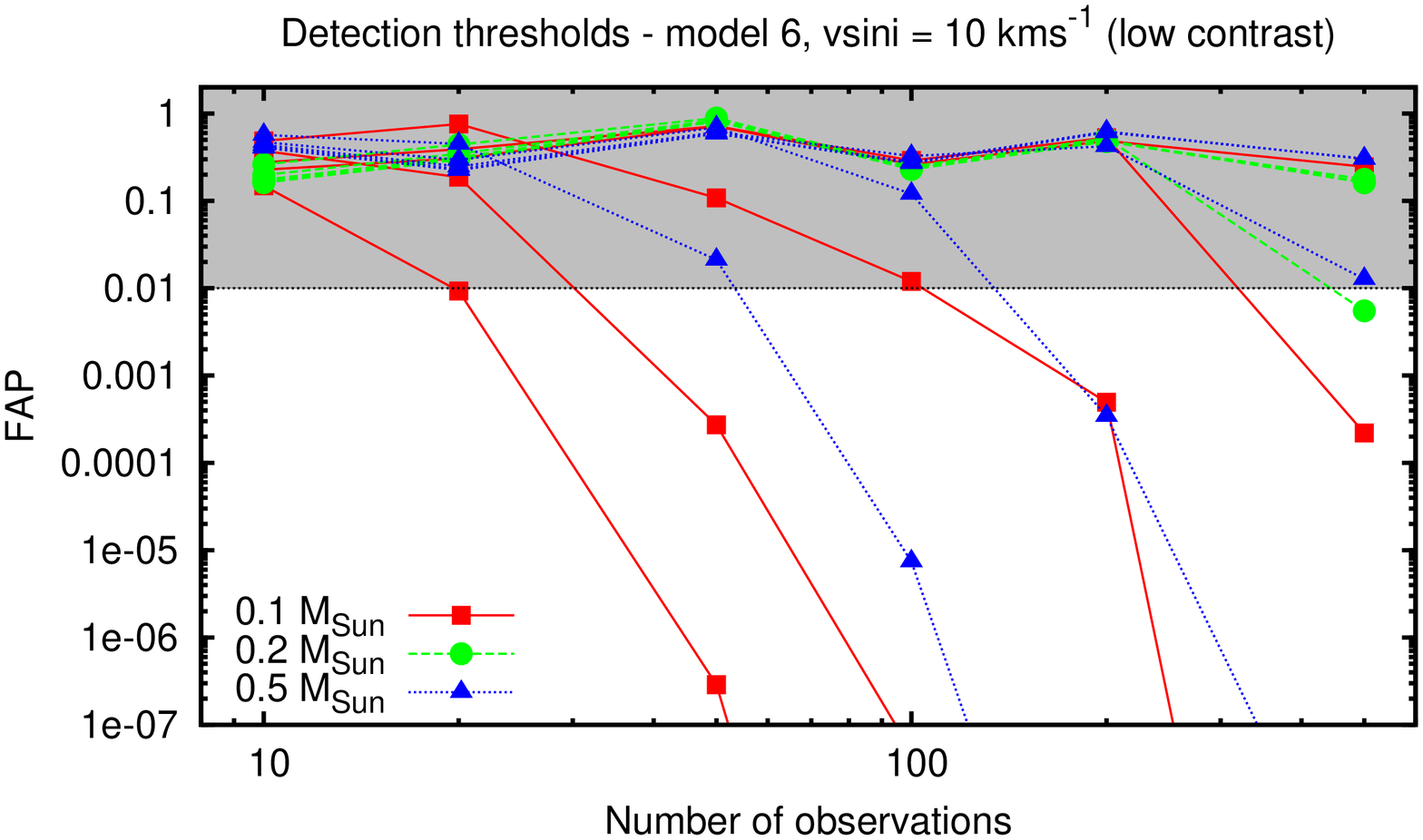}
 & 
\includegraphics[bbllx=80,bblly=100,bburx=510,bbury=750,width=54mm,height=85mm,angle=270]{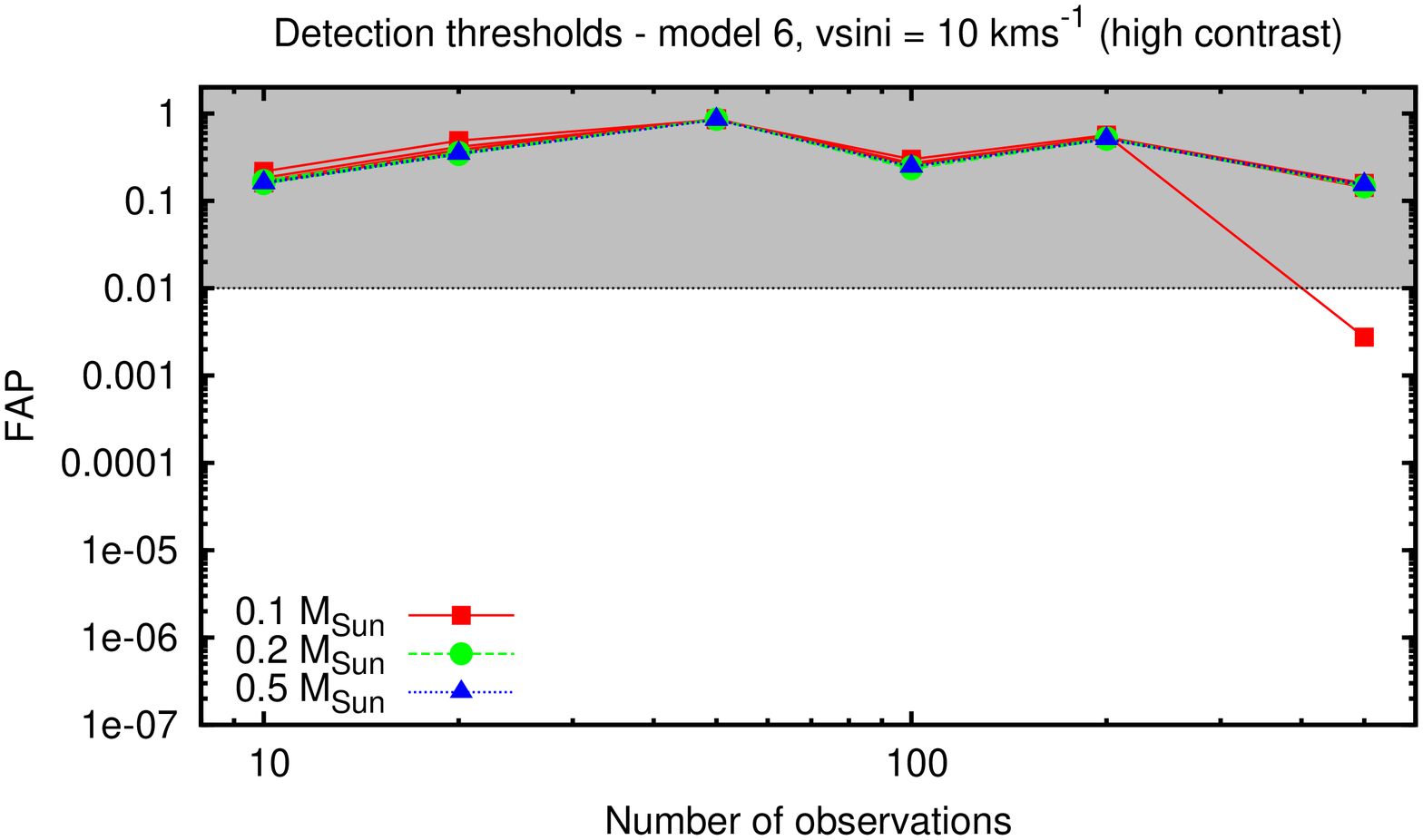} \\

\end{tabular}
\end{center}
\vspace{2mm} 
\caption{As for Fig. 5 for activity models 3 (top) and 6 (bottom) with
\vsini\ = 10 \kms (instrumental precision = 3 \kms).}
\protect\label{fig:planet_sim3}
\end{figure*}

\begin{figure*}
\begin{center}
\begin{tabular}{cc}

\includegraphics[bbllx=80,bblly=100,bburx=510,bbury=750,width=54mm,height=85mm,angle=270]{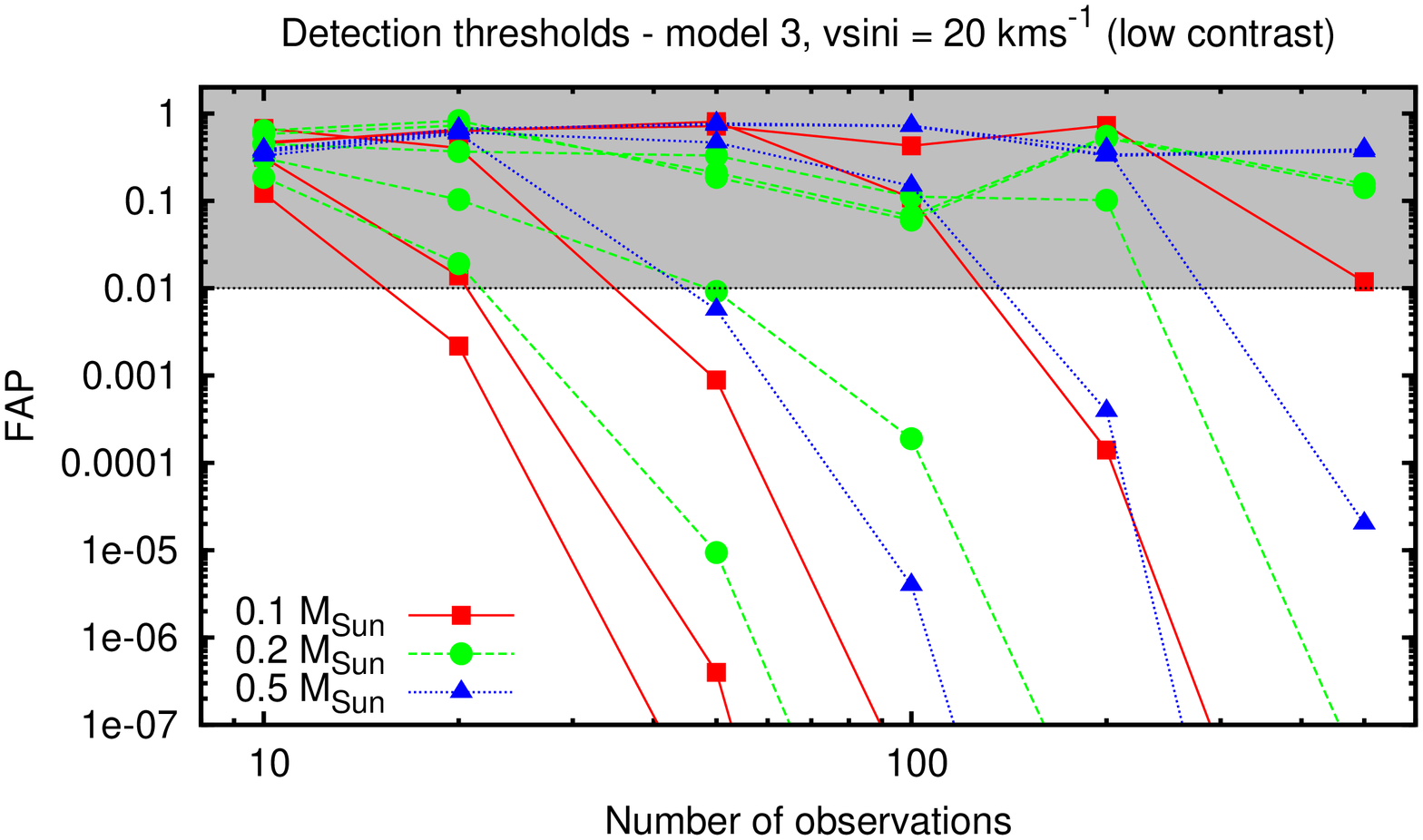}
 & 
\includegraphics[bbllx=80,bblly=100,bburx=510,bbury=750,width=54mm,height=85mm,angle=270]{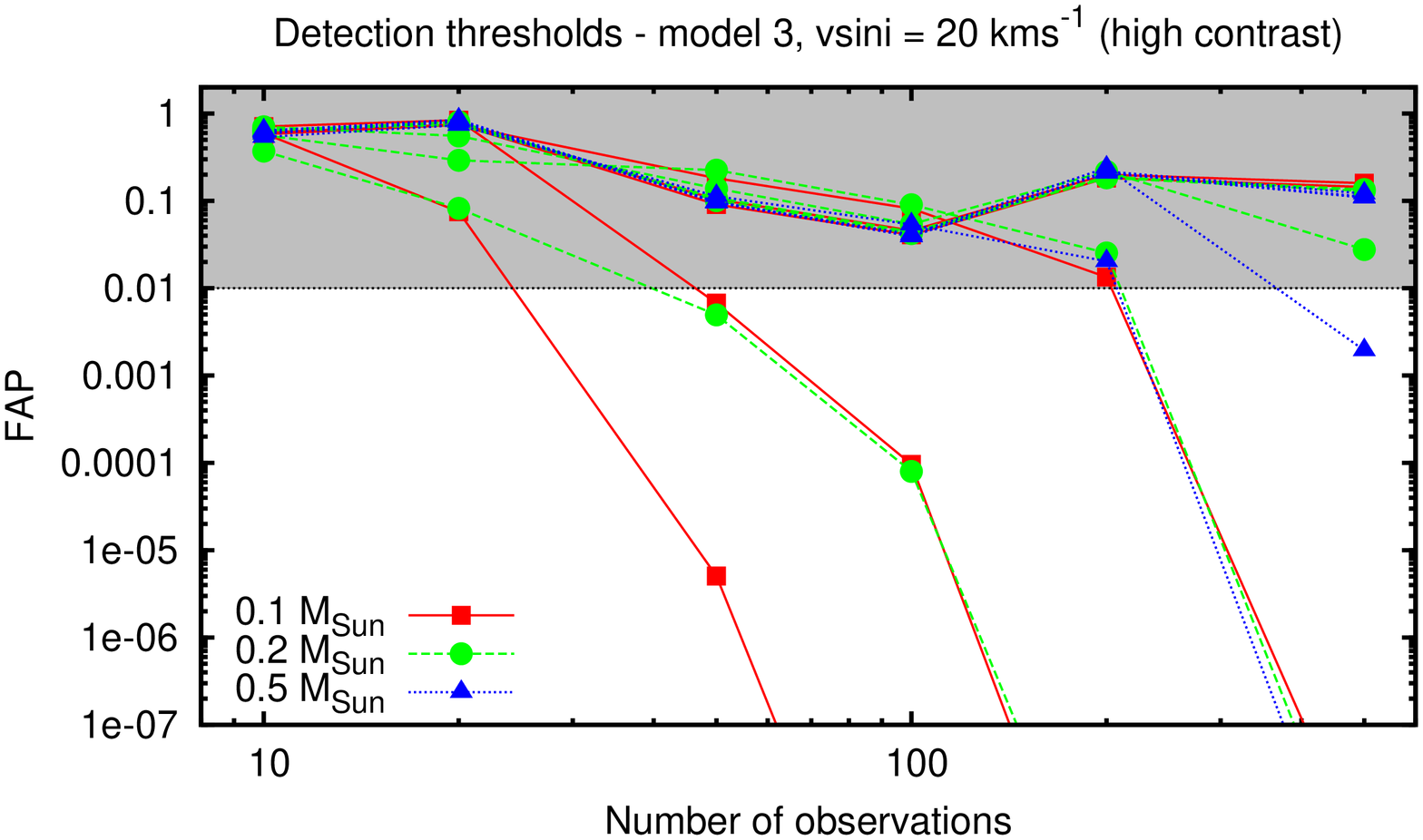} \\
\vspace{-6mm} \\
\includegraphics[bbllx=80,bblly=100,bburx=510,bbury=750,width=54mm,height=85mm,angle=270]{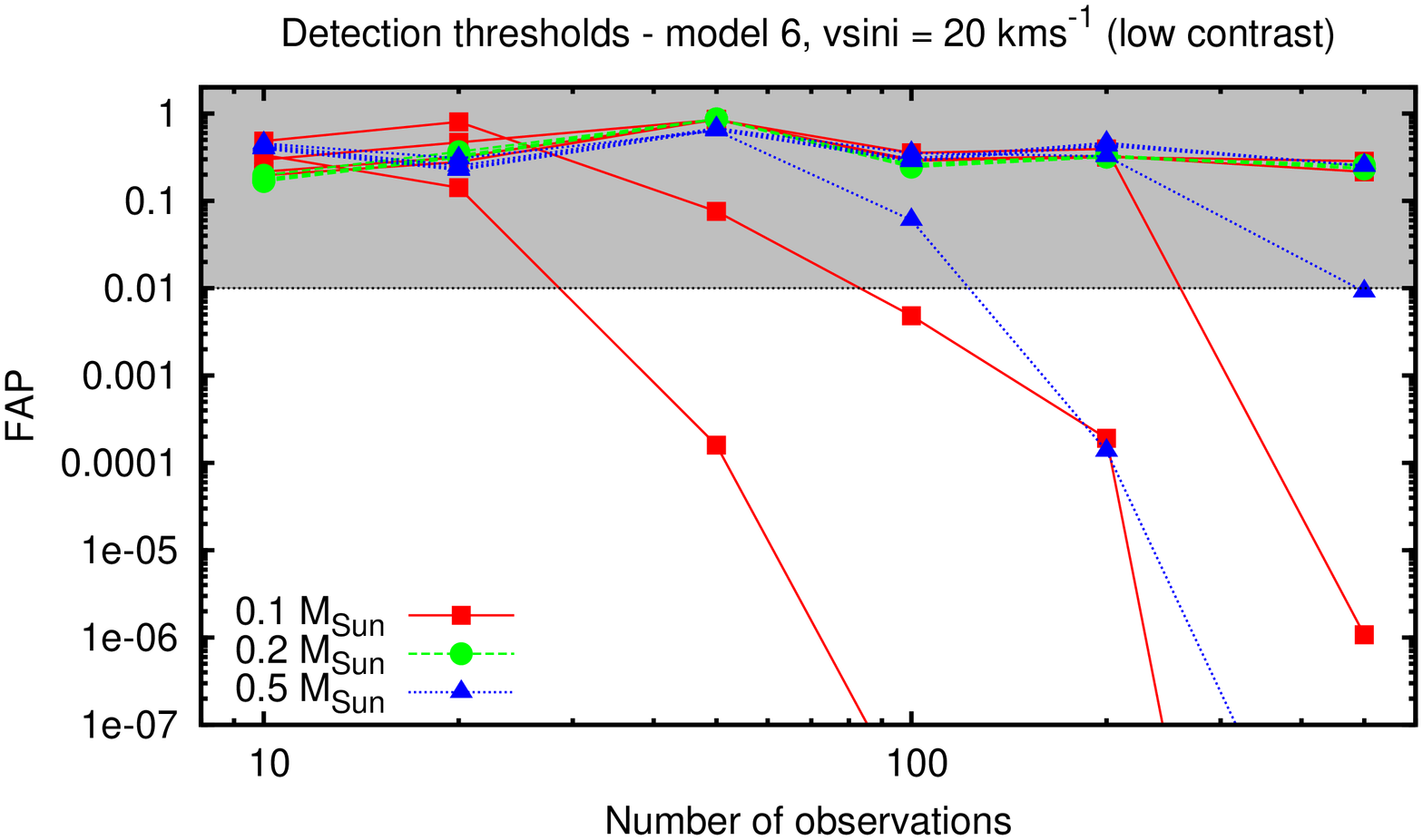}
 & 
\includegraphics[bbllx=80,bblly=100,bburx=510,bbury=750,width=54mm,height=85mm,angle=270]{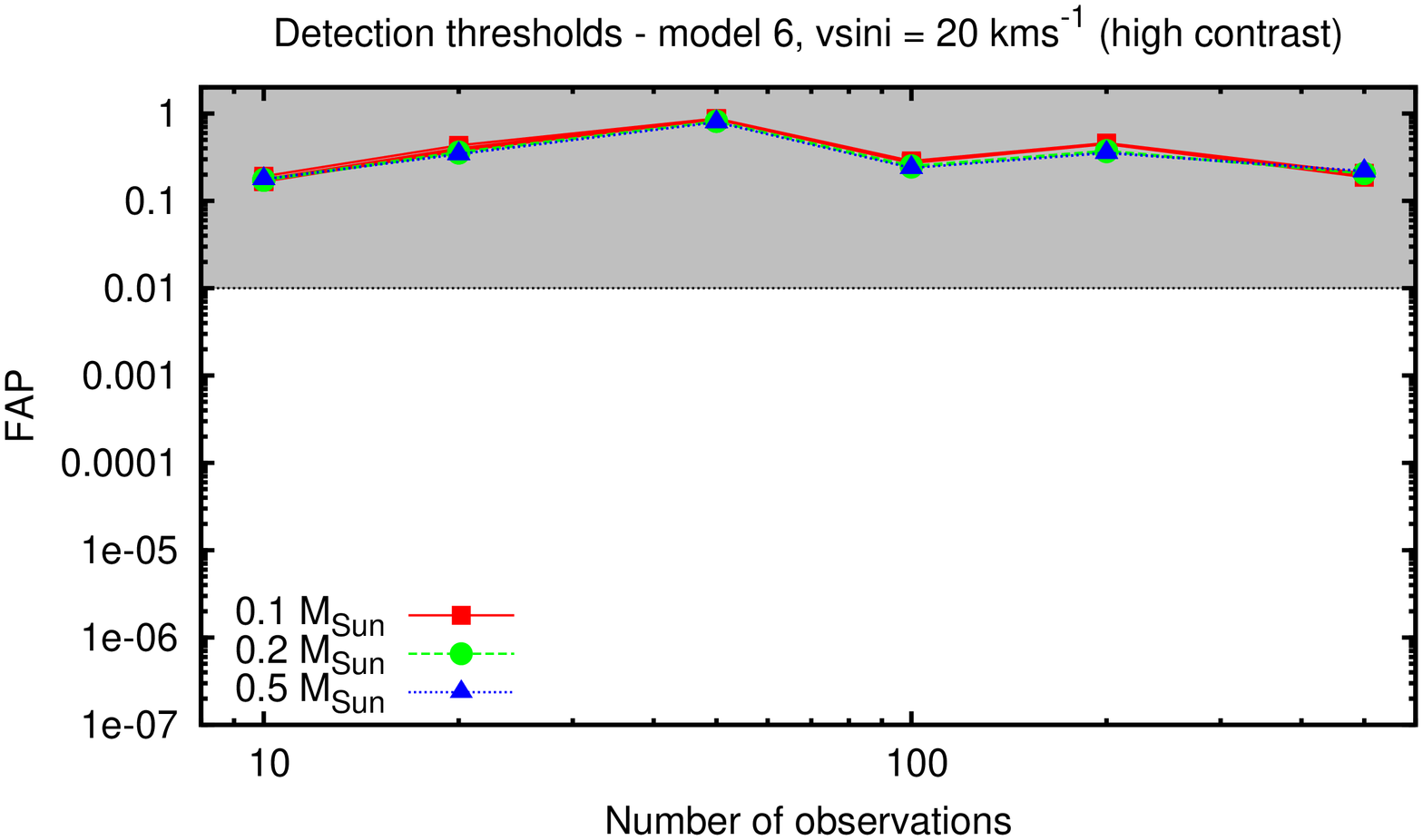} \\

\end{tabular}
\end{center}   
\vspace{5mm} 
\caption{As for Fig. 5 for activity models 3 (top) and 6 (bottom)
with \vsini\ = 20 \kms (instrumental precision = 6 \kms).}
\protect\label{fig:planet_sim4}
\end{figure*}

\section{Detection thresholds for M dwarf stars with starspots}
\protect\label{section:detection}

In this section, we simulate detection thresholds for low-mass planets orbiting
 at the centre of the habitable zone of M dwarf stars. We use the starspot
models from the preceding section to derive activity induced jitter. Since the
parameter space for characterising detection thresholds is so large, we have
chosen to fix a number of parameters as in previous sections. We have simulated
the detection thresholds for { 1, 2, 5, 10 \& 20 \mearth\ planets that orbit
in the {\em habitable zones} of 0.1, 0.2 \& 0.5 \msun\ stars. Table 2 lists
the periods of the planets for each stellar mass and the corresponding stellar
radial velocity amplitudes, $K_*$, for each planet. In this section we have
simulated all stellar and planetary orbit inclinations with $i = 90$\degs, and
used a range of \vsini\ values as discussed in \S 3.4. 

For each star/planet combination, we generate radial velocity points for a
range of observation epochs. We simulate 10, 20, 50, 100, 200 \& 500
epochs, with { one observation made every night} for simplicity.
Jitter from the two sources (i.e. starspots and instrumental/measurement
precision) is then added to the planetary radial velocities. The starspot
models 1\,-\,6 are used to add the stellar activity jitter to each planetary radial
velocity point. This is achieved by sampling, at a random observation
phase, the line profile that arises from the particular starspot model we are
interested in. We use model estimates to approximate the
instrumental/measurement precision. This is an important further consideration,
since \vsini\ affects the best precision that can be achieved. For an assumed Y
band S/N ratio of 90\,-\,100 and R$\sim$70,000, Fig. 7 of \cite{reiners10rvs}
indicates the appropriate accuracy that may be achieved in the Y band for a
3000 K atmosphere. This varies from $\sim$ 1.5 \ms\ at \vsini\ = 2 \kms\ to $\sim$
11 \ms\ at \vsini\ = 50 \kms. An example of a simulated planet induced stellar
RV signal is shown in Fig. \ref{fig:example_planet} before (solid/red) and after
(dashed/green) jitter is added. This planet is detected with FAP $<$ 0.01 in
Fig. \ref{fig:planet_sim3} (upper right panel, green square at 100 epochs).

}

We then carry out a Lomb Scargle periodogram analysis \citep{press92} on each
radial velocity curve in an attempt to recover the planetary signature. Although
this is adequate in our simulations, we note that we have not included the
effects of orbital eccentricity (i.e. $e=0$ in all simulations), which will
modify our detection thresholds \citep{ford05} in extreme cases. The results of
our periodogram analyses are plotted in Figs.
\ref{fig:planet_sim1}\,-\,\ref{fig:planet_sim4} as detection thresholds. { We
have selected to illustrate a range of scenarios that represent varying
\vsini\ and activity levels, following the preceding discussion. 


}

The false alarm probability (FAP) is plotted against the total number of
observations (see caption of Fig. \ref{fig:planet_sim1} for full details). The
left hand plots show the results for the cases where \hbox{$T_s = T_p - 200$ K} (low
contrast) while the right hand plots show the cases for which $T_s/T_p$ = 0.65
(high contrast). As expected, our ability to detect a planetary signature
decreases with increasing { \vsini\ and stellar activity level}. 

{
Fig. \ref{fig:planet_sim1} represents our low \vsini\ and activity (model 2 -
active solar analogue) level. Since both models 1 \& 2 exhibit significantly
$<$ 1 \ms\ { starspot jitter} (see Fig. 3) with \vsini\ = 2 \kms, we illustrate the
results for model 2 only since \cite{reiners10rvs} predict that the highest
precision achievable in the Y band with \vsini\ = 2 \kms\ is $\sim$1.5 \ms. The
precision therefore dominates the noise in this scenario rather than the
starspot jitter. As few as 20\,-30 epochs of observations are required to detect
$\leq$2 \mearth\ planet orbiting a 0.1 \msun\ star, while 50 epochs are
required to detect a 1 \mearth\ planet. However for a 0.5 \msun\ star, 500
epochs are required to detect a 1 \mearth\ planet. Obtaining such a large
number of observations is unlikely to be feasible, at least with limited
telescope allocations and a sufficiently large target sample. There is
little difference between the low and high contrast scenarios, again because
all starspot induced jitter is well below the achievable precision level.

Fig. \ref{fig:planet_sim2} illustrates results for \vsini\ = 5 \kms\ and two
different activity models (model 2 \& model 4 - with up to 9 per cent spot
filling). In this instance, for model 2, the changes are slight compared with
the preceding case where \vsini\ = 2 \kms. This observation illustrates a
further important limitation to precision, namely the instrumental resolution.
A greater degree of spottedness is required (i.e. model 4 cases in Fig.
\ref{fig:planet_sim2}) before a noticeable change in detection thresholds begins
to take place. Here, while 80 epochs in the low contrast case will enable 1
\mearth\ planets to be detected, $\sim$500 epochs are required to detect 1
\mearth\ planets in the high contrast scenario. \cite{jenkins09mdwarfs} report a
median \vsini\ for 0.5 \msun\ and 0.2 \msun\ stars that most closely match the
\vsini\ = 5 \kms\ used in Fig. \ref{fig:planet_sim2}. This simulation may
therefore be deemed to represent the detectability of earth-mass to
few-earth-mass planets around {\em average} 0.5 \msun\ and 0.2 \msun\ stars.

In Fig. \ref{fig:planet_sim3}, we show the expected detectability of planets
around stars with \vsini\ = 10 \kms. This rotation velocity is the median for
later M spectral types, such as the 0.1 \msun\ star. By 10 \kms, the less active
model 3 scenario still enables \hbox{2 \mearth}\ planets orbiting 0.1 \msun\ 
stars to be detected in the low contrast case, whereas 5 \mearth\ planets may be
detected in the same number of epochs for the high contrast case. Only planets
with mass $\geq$ 10 \mearth\ may be detected with $<$ 100 epochs of observations
for the high contrast case. With a highly active star (model 6), only one planet
(20 \mearth) remains detectable in the high contrast regime, albeit requiring
500 epochs of observations. For model 6, the starspot jitter contribution is so
large (compared with the instrumental/measurement precision) that it dominates
the noise contribution. We also noted in \S 3.3 (see also Fig. { 3}) that the
jitter is greatest for the 0.2 \msun\ model with low spot contrast spots. As a
result, in this scenario, our model predicts that detection of a planet orbiting
a 0.5 \msun\ star becomes easier than for a planet orbiting a 0.2 \msun\ star.

The global trend seen in Figs. 5\,-\,7 continues in Fig. \ref{fig:planet_sim4},
where only the most massive planets may be detected for very active stars. If
the spot filling factors of up to 50 per cent, reported at earlier spectral
types for stars rotating with $\sim$ 20 \kms\ \citep{oneal98tio} are to be found
in M dwarfs, then a lower planet detection limit of order 1 Neptune mass may be
expected. By 50 \kms (not shown), only 20 \mearth\ planets may be detected in orbit around
a 0.1 \msun\ star with low contrast spots (model 6) in less than 100 epochs of
observations (for model 4, only $\geq$10 \mearth\ planets may similarly be
detected in $\leq 100$ epochs).

{
\subsection{The effects of activity and observing strategy on planet detectability}
\protect\label{section:detectability}

We have demonstrated how the relative contrast ratio, $F_p/F_s$ is an important factor that determines the detectability of planets, particularly when spot activity dominates the jitter. However, it must also be realised that the observing strategy simulated in this paper will determine the detectability of planets. In Particular, the orbital period of the planet may be either somewhat shorter, or longer than the span of the observations. Table 2 indicates that the orbital periods of the simulated habitable zone planets are $\sim$ 5, 13 \& 36 days for 0.1, 0.2 \& 0.5 \msun stars respectively. Hence, detection of a periodic signal on timescales shorter than the period is less likely. In other words, it should be easier to detect habitable zone planets orbiting lower mass stars owing to the shorter periods. Since we only simulate a minimum of 10 epochs, this mostly applies only to the 36 day period experienced by a planet orbiting a 0.5 \msun\ star. It can be seen from Figs. 5\,-\,8 that no planets are detected in orbit about a 0.5 \msun\ star until over half the period has been sampled. Therefore if trageting such stars for habitable zone planets, a different observing strategy would be needed to minimise the number of observation epochs. In reality, a radial velocity survey would wish to search for planets with a range of orbital radii and periods, especially as shorter period planets are more likely to be detected.

The effect of activity on the number of epochs required to detect a planet can also be assessed by completely removing the starspot jitter from all simulations. In our low activity simulation (vsini = 2 \kms\, instrumental precision of 1.5 \ms\ and Model 2 starspot coverage), removing the starspot activity does not have any significant effect on the detection curves plotted in Fig. 5. This has already been discussed in the preceding section and is a consequence of the instrumental precision at this rotation velocity (and implied instrumental precision) dominating the jitter. The effect of removing spot activity from the simulations presented in Figs. 6\,-\,8 however leads to significant changes in the number of epochs necessary for detection, as might be expected. In our most extreme scenario simulated in Fig. 8 (high contrast), although no planets are detected, complete removal of the starspot jitter leaves a 6 \ms\ precision floor due to the rotation velocity of \vsini\ = 20 \kms. Here, planets of 5\,-\,20 \mearth\ may then be detected with 300\,-\,40 epochs respectively.

}
}

\section{Summary \& Conclusions}
\protect\label{section:discussion}

{ We have used more realistic starspot models than previous studies to
determine the effect of activity induced jitter in precision radial velocity
studies of M dwarf stars.} We have demonstrated that with only several tens of
epochs, habitable zone earth-mass planets can be detected around low-activity
stars. Since the contrast ratio between photosphere and spots is uncertain, we
opted to simulate two extreme cases. In reality, the true
contrast ratio likely falls between the two extreme cases simulated.
\cite{rockenfeller06mdwarfs} required temperature differences of only
a few hundred K in order to fit their lightcurves of M5V \& M9V stars.
However, their modelling assumed only a single spot was present. A more
uniformly spotted surface, especially if highly spotted could produce a similar
lightcurve but would be expected to require a higher contrast between
photosphere and spot to achieve the same photometric amplitude. 

The factors that determine the starspot induced jitter have important
consequences for the estimated detection threshold limits. { In agreement
with other studies \citep{reiners10rvs}} we showed (Fig. 1)
that once the contrast ratio between spot and photosphere is sufficiently high,
the improvement in jitter, when moving from V-band to Y-band, is less
impressive. In this case, the spots are simply dark enough that the radial
jitter does not show such a strong decrease with increasing
wavelength. Conversely, the decrease in jitter as a function of increasing
wavelength is much more pronounced at lower contrast ratios. At low contrast,
the exact normalised line strengths and relative line strengths of the
photosphere and spot play an important role in determining the exact relative
jitter at different wavelengths. This additional factor leads to the
{ irregularity in relative RV amplitudes as a function of increasing 
$T_p$, as shown in Fig. 2 (see \S 3.2).} { Although we do
not show the full jitter amplitude ratio for other photometric bands, Fig. 2 (top) for $T_p = 3250$ K
indicates that the relationship between wavelength and jitter is not as straightforward
as a simple blackbody model might suggest. Further estimates of the wavelength dependence 
of jitter may be obtained by observation of the results of
\cite{reiners10rvs} who plot RV amplitude as a function of wavelength for the
5000 \AA\,-\,18000 \AA\ range (their Fig. 12). At longer wavelengths than Y band, equivalent width effects at low photosphere/spot contrast apart, there is little further gain in precision. A further important consideration is the number of lines available for cross-correlation, which may be fewer at longer wavelengths.}

{ We have assumed that activity scales with rotation \citep{browning10} so
that
less spotted stars are slow rotators and more spotted stars are fast rotators.
However, as discussed in \S \ref{section:observedspots},
\cite{reiners10activity} { find that the relation between rotation and 
activity is weaker in M dwarfs.} Knowledge of the true spottedness of
these stars is clearly vital for an accurate estimation of the detectability
limits for low mass planets orbiting the lowest mass M dwarfs.  In addition,
spot coverage derived from
other spectroscopic methods using temperature
sensitive (TiO) lines \citep{oneal98tio} indicate that there may be large
discrepancies with results from Doppler images. For active stars, up to $\sim$
50 per cent spot coverage has been derived { using the TiO method}, while Doppler images of similar
stars typically derive $< 10$ per cent coverage. { The difference between these methods is
that the TiO procedure derives an unresolved mean spot coverage while the Doppler imaging results 
are limited by the amount of resolvable information in the broadened rotation profile of the star.}
It seems likely that active
stars are therefore more spotted than Doppler images show, { and} that they
possibly exhibit smaller spots than are derived via this method. In
summary, high spectral resolution Doppler imaging surveys (in the red-optical or
infrared), possibly combined with other global spot coverage surveys are needed
for a more accurate picture of spot patterns on mid-late M dwarfs.}

It is clear that there are many physical factors that will determine the true
detection threshold for planets orbiting M dwarfs; { we have only
simulated planets in circular orbits for instance. \cite{kennedy08}
simulated planet scattering and found that for low mass stars only, planets with
long circularisation times on eccentric orbits could form. Although we have not
simulated the effects of eccentricity in full, we have already noted that in
extreme cases, detection thresholds will be raised and more observations will be
required \citep{ford05}.  We find that for longitude of periastron, $\omega =
0$\degs, \vsini\ = 10 \kms\ (instrumental precision = 3 \ms), the number of
nights required to make a detection increases by approximately 2\,-\,2.5 times
for eccentricity, $e = 0.5$ and by $\sim 5$ times for $e = 0.9$.} 

In this paper, we have shown that earth-mass or near earth-mass planets in the
habitable zones of late-mid M dwarfs can be detected when the number of
starspots matches those seen on the Sun at extremes of activity. { Moderate
rotation quickly increases the number of observations that are required to make
a detection such that by the time \vsini\ = 10 \kms, $\geq 100$ observations
are required to detect planets with masses $\leq$ 5 \mearth.} Variations in
\vsini\ and contrast ratio are therefore important factors in particular in
determining the detection thresholds. Given the evidence for a general increase
in \vsini\ with decreasing mass
\citep{delfosse98mdwarfs,mohanty03activity,jenkins09mdwarfs}
among M dwarfs, the balance of these factors may be important. Obtaining
estimates of starspot coverage from further modelling and photometric
observations will provide vital information that will enable a link
between starspots and more traditional chromospheric activity indicators
\citep{delfosse98mdwarfs,mohanty03activity,reiners09activity,browning10} to be
made.

\section*{Acknowledgments}

We would like to thank David Parker (Monk's Walk School, Herts. UK) and
Robert Sanders (Richard Hale School, Herts. UK) for their contributions to
this work. { We would also like to thank the referee,
Ansgar Reiners, for constructive comments that have resulted in an improved
final version of the manuscript.} SVJ currently acknowledges support from De Nederlandse Organisatie
voor Wetenschappelijk Onderzoek (NWO).


\protect\label{lastpage}
\end{document}